\documentclass[usenatbib]{mn2e}
\usepackage{psfig}

\def\apj{ApJ}
\def\apjl{ApJL}
\def\mnras{MNRAS}
\def\pasp{PASP}

\def\araa{ARAA}
\def\aap{A\&A}
\def\aj{AJ}
\def\apjs{ApJS}

\def\nat{Nature}

\def\gs{\mathrel{\raise0.35ex\hbox{$\scriptstyle >$}\kern-0.6em\lower0.40ex\hbox{{$\scriptstyle \sim$}}}}
\def\ls{\mathrel{\raise0.35ex\hbox{$\scriptstyle <$}\kern-0.6em\lower0.40ex\hbox{{$\scriptstyle \sim$}}}}

\def\Wm2{\,\hbox{W}\,\hbox{m}^{-2}}
\def\gsim{\mathrel{\raise0.35ex\hbox{$\scriptstyle >$}\kern-0.6em\lower0.40ex\hbox{{$\scriptstyle \sim$}}}}
\def\lsim{\mathrel{\raise0.35ex\hbox{$\scriptstyle <$}\kern-0.6em\lower0.40ex\hbox{{$\scriptstyle \sim$}}}}

\begin{document}

\title[The Dynamics and Metallicity Gradients of Star-Forming Galaxies
  at $z$\,=\,0.84--2.23]{The properties of the star-forming
  interstellar medium at $z$\,=\,0.84--2.23 from HiZELS: Mapping the
  internal dynamics and metallicity gradients in high-redshift disk
  galaxies}

\author[Swinbank et al.]
{\parbox[h]{\textwidth}{
A.\,M.\ Swinbank,$^{\, 1,*}$
D.\, Sobral,$^{\, 2}$
Ian Smail,$^{\, 1}$
J.\,E.\ Geach,$^{\, 3}$
P.\,N.\ Best,$^{4}$
I.\,G\ McCarthy,$^{5}$
R.\,A\ Crain,$^{2}$
\& T.\, Theuns$^{1,6}$
}
\vspace*{6pt} \\
$^1$Institute for Computational Cosmology, Durham University, South Road, Durham, DH1 3LE, UK \\
$^2$Leiden Observatory, Leiden University, PO Box 9513, 2300 RA Leiden, the Netherlands\\
$^3$Department of Physics, McGill University, Ernest Rutherford Building, 3600 Rue University, Montreal, Quebec H3A 2T8, Canada\\
$^4$SUPA, Institute for Astronomy, University of Edinburgh, Edinburgh, EH19 3HJ, UK \\
$^5$School of Physics and Astronomy, University of Birmingham, Edgbaston, Birmingham B15 2TT, UK\\
$^6$University of Antwerp, Campus Groenenborger, Groenenborgerlaan 171, B-2020 Antwerp, Belgium\\
$^*$Email: a.m.swinbank@durham.ac.uk \\
}

\maketitle

\begin{abstract}
  We present adaptive optics assisted, spatially resolved spectroscopy
  of a sample of nine H$\alpha$-selected galaxies at $z$\,=\,0.84--2.23
  drawn from the HiZELS narrow-band survey.  These galaxies have
  star-formation rates of 1--27\,M$_{\odot}$\,yr$^{-1}$ and are
  therefore representative of the typical high-redshift star-forming
  population.  Our $\sim$kpc-scale resolution observations show that
  approximately half of the sample have dynamics suggesting that the
  ionised gas is in large, rotating disks.  We model their velocity
  fields to infer the inclination-corrected, asymptotic rotational
  velocities.  We use the absolute $B$-band magnitudes and stellar
  masses to investigate the evolution of the $B$-band and stellar mass
  Tully-Fisher relationships.  By combining our sample with a number of
  similar measurements from the literature, we show that, at fixed
  circular velocity, the stellar mass of star-forming galaxies has
  increased by a factor 2.5 between $z$\,=\,2 and $z$\,=\,0, whilst the
  rest-frame $B$-band luminosity has decreased by a factor $\sim$\,6
  over the same period.  Together, these demonstrate a change in
  mass-to-light ratio in the $B$-band of $\Delta$(M\,/\,L$_{\rm
    B}$)\,/\,(M\,/\,L$_{\rm B}$)$_{\rm z=0}\sim$3.5 between $z$\,=\,1.5
  and $z$\,=\,0, with most of the evolution occuring below $z$\,=\,1.
  We also use the spatial variation of [N{\sc ii}]\,/\,H$\alpha$ to
  show that the metallicity of the ionised gas in these galaxies
  declines monotonically with galactocentric radius, with an average
  $\Delta$\,log(O\,/\,H)\,/\,$\Delta$\,R\,=\,$-$0.027\,$\pm$\,0.005\,dex\,kpc$^{-1}$.
  This gradient is consistent with predictions for high-redshift disk
  galaxies from cosmologically based hydrodynamic simulations.
\end{abstract}

\begin{keywords}
  galaxies: evolution -- galaxies: formation -- galaxies: high-redshift
\end{keywords}

\section{Introduction}
\label{sec:intro}

Numerical simulations indicate that the majority of massive galaxies at
$z$\,=\,1--3, the epoch when galaxies were most rapidly growing their
stellar mass, are continuously fed by gas which promotes and maintains
star formation \citep{Bournaud09,Dekel09,VandeVoort11,VandeVoort12}.
The accretion of gas from the halo, through minor mergers or by cold
streams from the inter-galactic medium appears to dominate over that
from major mergers and suggests that the star formation within the
inter-stellar medium (ISM) of high-redshift galaxies are driven by
internal dynamical processes \citep{Keres05}.  The observational
challenge is now to quantitatively measure the internal properties
(e.g.\ gas surface density, disk scaling relations, chemical
abundances, distribution and intensity of star-formation) and so test
whether the prescriptions developed to describe star-formation
processes within disks at $z$\,=\,0 are applicable in the rapidly
evolving ISM of gas rich galaxies at high-redshift
\citep{Krumholz05,Hopkins12a}.

High-spatial resolution observations of star-forming galaxies around
$z\sim$\,1--2 have shown that a large fraction of the population have
their ionised gas in large, rotating disks
\citep[e.g.\ ][]{Genzel06,ForsterSchreiber06,ForsterSchreiber09,Stark08,Jones10,Wisnioski11}.
In contrast to low-redshift, these disks have high gas fractions
($f_{gas}$\,=\,20--80\%) and so are turbulent, with much higher
velocity dispersions given their rotational velocity
($\sigma$\,=\,30--100\,km\,s$^{-1}$, v$_{\rm
  rot}$\,/\,$\sigma\sim$\,0.2--1) compared to the thin disks of local
spirals \citep{Tacconi10,Daddi10,Geach11,Swinbank11}.

If low- and high- redshift disks can be linked in a single evolutionary
model, then the redshift evolution of disk scaling relations
(i.e.\ between size, luminosity, rotational speed and gas fraction)
providing a diagnostic of the relationship between the recent- and past
averaged- star-formation with the dark halo and the dominant mode of
assembly.

The pioneering work of \citet{Vogt96} and \citet{Vogt99} provided the
first systematic study of the evolution of the stellar luminosity
(M$_{\rm B}$) versus circular velocity (Tully-Fisher;
\citealt{TullyFisher}) relation, deriving an increase in luminosity at
a fixed circular velocity of $\Delta$L$_{\rm B}$\,=\,1.7\,$\times$ up
to $z\sim$\,1, reflecting the change in star formation efficiency over
this period (see also \citealt{Bamford05,FernandezLorenzo09} and
\citealt{Miller11}).  Since the $B$-band luminosity is sensitive to
recent star formation, attempts have also been made to measure the
evolution of the stellar mass (M$_{\star}$) Tully-Fisher relation which
reflects the relation between the past-average star-formation history
and halo mass.  Although the current samples sizes are drawn from a
heterogeneous mix of populations, evidence is accumulating that there is
only modest evolution in the stellar-mass Tully-Fisher relation to
$z\sim$\,1.5, with stronger evolution above $z\sim$\,1.5
\citep{Miller12}.  Theoretical mode of galaxy formation have struggled
to reproduce the zero-point and slope of the Tully-Fisher relation,
tending to produce disks that rotate too fast given their luminosities
\citep[e.g.][]{Mo98,Benson03,Dutton11}, although some of this
discrepancy has been alleviated by improving the recipes for starburst
driven feedback
\citep[e.g.\ ][]{Sales10,Tonini11,Piontek11,Scannapieco11,Desmond12,McCarthy12}.

A second potential observational consequence of differences in fueling
mechanisms for disk galaxies is their chemical abundance distributions.
The primary indicator of chemical evolution is typically traced by
Oxygen as its relative abundance surpasses all elements heavier than
Helium and it is present almost entirely in the gas phase
\citep{Snow96,Savage96}.  

In the thin disk of the Milky-Way and other
local spirals, there is a negative radial metallicity gradient
\citep[e.g.\ ][]{Searle71,Shields74,McCall85,VilaCostas92,Zaritsky94,Ferguson98,Zee98}.
In contrast, the thick disk of the Milky Way displays no radial
abundance gradient \citep{Gilmore95,Bell96,Robin96,Edvardsson93}.
Thick disks are ubiquitous in spiral galaxies, and typically contain
10--25\% of their baryonic mass, yet the thick disk of the Milky-Way
contains no stars younger than 8\,Gyr \citep{Gilmore85,Reddy06c}.
Thick disks are comparable in many physical properties to the
apparently turbulent disks of high-redshift galaxies, and so
determining the early form of their abundance gradient (positive or
negative) will provide a key indication of how quickly enrichment
proceeds in inner and outer regions, as well as the interplay between
star formation, clump migration and gas accretion from the halo or IGM \citep{Solway12}.

Indeed, if the majority of the gas accretion in high-redshift galaxies
is via accretion from the IGM along filaments which intersect and
deposit pristine material onto the galaxy disk at radii of 10--20\,kpc,
then the inner disks of galaxies should be enriched by star-formation
and supernovae whilst the the outer-disk continually diluted by
pristine gas, leaving strong negative abundance gradients
\citep{Dekel09} which will flatten as the gas accretion from the IGM
becomes less efficient, and the gas redistributed, at lower redshift.

To gain a complete census of how galaxies assemble the bulk of their
stellar mass, it is clearly important to study the dynamics, chemical
properties and internal star-formation processes in a well-selected
sample of high-redshift galaxies.  To this end, we have combined
panoramic (degree-scale) imaging with near-infrared narrow-band
filters, to carry out the High-Z Emission Line Survey (HiZELS) survey,
targeting H$\alpha$ emitting galaxies in four precise ($\delta
z$\,=\,0.03) redshift slices: $z$\,=\,0.40, 0.84, 1.47 \& 2.23
\citep{Geach08,Sobral09,Sobral10,Sobral11,Sobral12a,Sobral12b}.  This
survey provides a large, luminosity-limited sample of {\it identically}
selected H$\alpha$ emitters at epochs spanning the apparent peak of the
cosmic star-formation rate density, and provides a powerful resource
for studying the properties of representative galaxies
(i.e.\ star-formation rates $\sim$10's\,M$_\odot$/yr) which will likely
evolve into $\sim L^{\star}$ galaxies by $z$\,=\,0, but are seen at a
time when they are assembling the bulk of their stellar mass, and thus
at a critical stage in their evolutionary history.  One of the key
benefits of selecting a high-redshift galaxy population using H$\alpha$
alone is the simplicity in comparison to similar luminosity-limited
samples of star-forming galaxies in the local Universe.  Together,
these comparison samples can be used to search for differences in the
physical properties of the ISM that potentially provide insight into
the factors influencing star formation at high redshift.

In this paper, we present adaptive optical assisted ($\sim$kpc scale)
integral field spectroscopy with SINFONI of nine star-forming galaxies
selected from the HiZELS survey in the redshift range
$z$\,=\,0.84--2.23 -- the SINFONI-HiZELS (SHiZELS) survey.  We use
these data to investigate the dynamical properties of the galaxies, the
evolution of the luminosity and stellar mass scaling relations (through
the Tully-Fisher relation), and the star formation and enrichment
within their ISM.  We use a cosmology with $\Omega_{\Lambda}$\,=\,0.73,
$\Omega_{m}$\,=\,0.27, and H$_{0}$\,=\,72\,km\,s$^{-1}$\,Mpc$^{-1}$.
In this cosmology, at the median redshift of our survey, $z$\,=\,1.47,
a spatial resolution of 0.1$''$ corresponds to a physical scale of
0.8\,kpc.  All quoted magnitudes are on the AB system and we use a
Chabrier IMF \citep{Chabrier03}.

\begin{figure*}
  \centerline{\psfig{file=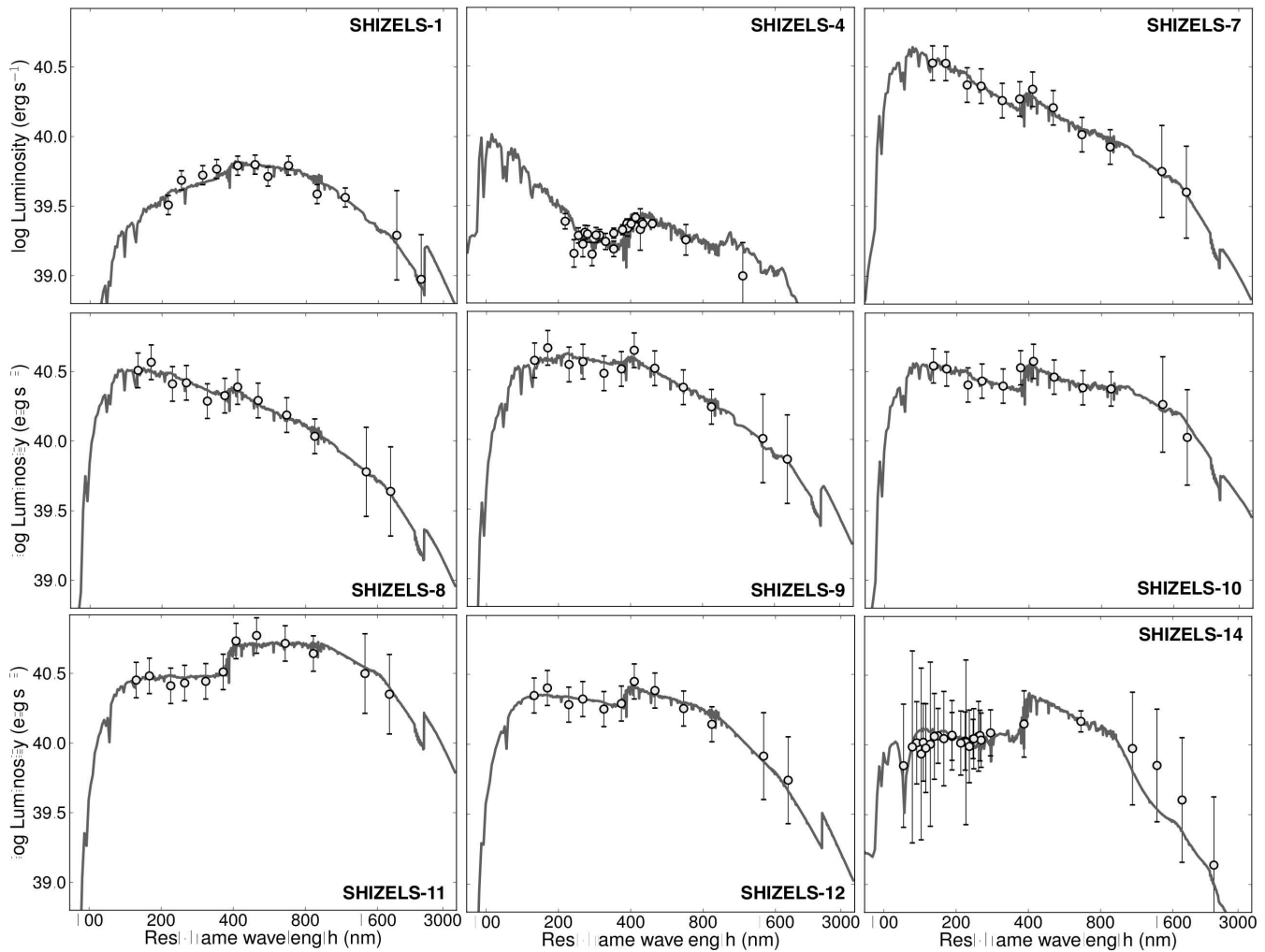,angle=0,width=7in}}
  \caption{Broad-band photometry and SEDs for the nine galaxies in our
    sample from rest-frame- UV to mid-infrared bands (spanning {\it
      GALEX} far-UV and near-UV bands to {\it Spitzer}/IRAC).  In all
    cases, we calculate the rest-frame spectral energy distribution
    (SED) and star-formation history and use this information to
    estimate the stellar masses (Table~2).  The SEDs are calculated
    using a $\chi^2$ fit using the \citet{Bruzual03} population
    synthesis models with a Chabrier IMF.  The SHiZELS galaxies sample
    a range of stellar masses from $\sim$10$^{9-11}$\,M$_{\odot}$ and
    specific star-formation rates 0.1--10\,Gyr$^{-1}$.}
\label{fig:SEDs}
\end{figure*}

\begin{table}
\begin{center}
{\tiny
{\centerline{\sc Table 1: Targets \& Observations}}
\begin{tabular}{lcccccc}
\hline
\noalign{\smallskip}
ID          & RA          & Dec          & $z_{\rm H\alpha}$   & $t_{\rm exp}$ & Strehl & EE      \\
            & (J2000)     & (J2000)      &                & $(\rm ks)$   &        & (0.1$''$) \\
            &             &              &                &          &        &          \\
\hline
SHiZELS-1   & 02\,18\,26.3 & $-$04\,47\,01.6 & 0.8425         & 9.8       & 15\%  & 11\%     \\
SHiZELS-4   & 10\,01\,55.3 & $+$02\,14\,02.6 & 0.8317         & 3.6       & 10\%  & 8\%     \\
\noalign{\smallskip}
SHiZELS-7   & 02\,17\,00.4 & $-$05\,01\,50.8 & 1.4550         & 13.4     & 27\%   & 25\% \\
SHiZELS-8   & 02\,18\,21.0 & $-$05\,19\,07.8 & 1.4608         & 9.8      & 21\%   & 27\% \\
SHiZELS-9   & 02\,17\,13.0 & $-$04\,54\,40.7 & 1.4625         & 9.8      & 24\%   & 27\% \\
SHiZELS-10  & 02\,17\,39.0 & $-$04\,44\,43.1 & 1.4471         & 9.8      & 20\%   & 25\% \\
SHiZELS-11  & 02\,18\,21.2 & $-$05\,02\,48.9 & 1.4858         & 3.6      & 13\%   & 32\% \\
SHiZELS-12  & 02\,19\,01.4 & $-$04\,58\,14.6 & 1.4676         & 9.8      & 20\%   & 30\% \\
\noalign{\smallskip}
SHiZELS-14  & 10\,00\,51.6 &  +02:33\,34.5  & 2.2418         & 12.0     & 12\%   & 26\% \\
\hline
\label{table:phot}
\end{tabular}
}
\caption{Notes: EE denotes the encircled energy within a radius of 0.1$''$.}
\end{center}
\end{table}

\begin{table*}
\begin{center}
{\small
{\centerline{\sc Table 2: Integrated Galaxy Properties}}
\begin{tabular}{lccccccccc}
\hline
\noalign{\smallskip}
ID          & f$_{H\alpha}$       & SFR$_{\rm H\alpha}$ & $r_{1/2}$        & [N{\sc ii}]\,/\,H$\alpha$ & M$_B$    &  M$_K$    & $\log_{\rm 10}$(M$_*$)  & E\,(B\,$-$\,V) & $\Delta$\,log(O\,/\,H)\,/\,$\Delta$\,R)\\
            & ($\times$10$^{-16}$ & (M$_{\odot}$/yr)   & (kpc)           &                       & (AB)       &  (AB)    & (M$_{\odot}$)       &          & (dex\,kpc$^{-1}$) \\                 
            & erg\,s$^{-1}$\,cm$^{-2}$) \\
\hline                                                                                                                                                                    
SHiZELS-1   & 1.0\,$\pm$\,0.1        & 2                 & 1.8\,$\pm$\,0.3     & 0.2\,$\pm$\,0.05   &  $-$18.70  & $-$21.54  & 10.03\,$\pm$\,0.15  & 0.4\,$\pm$\,0.1    & $-$0.037$_{-0.058}^{+0.030}$ \\
SHiZELS-4   & 0.7\,$\pm$\,0.1        & 1                 & 1.4\,$\pm$\,0.5     & 0.2\,$\pm$\,0.1    &  $-$20.68  & $-$21.49  &  9.74\,$\pm$\,0.12  & 0.0\,$\pm$\,0.2    & ... \\                         
SHiZELS-7   & 1.2\,$\pm$\,0.1        & 8                & 3.7\,$\pm$\,0.2     & 0.43\,$\pm$\,0.05  &  $-$21.12  & $-$22.15  &   9.81\,$\pm$\,0.28 & 0.2\,$\pm$\,0.2    & $-$0.019$_{-0.040}^{+0.019}$ \\
SHiZELS-8   & 1.1\,$\pm$\,0.1        & 7                & 3.1\,$\pm$\,0.3     & $<$\,0.1         &  $-$21.26  & $-$21.90  &  10.32\,$\pm$\,0.28 & 0.2\,$\pm$\,0.2    & +0.006$_{-0.004}^{+0.017}$   \\
SHiZELS-9   & 1.0\,$\pm$\,0.1        & 6                & 4.1\,$\pm$\,0.2     & 0.27\,$\pm$\,0.03  &  $-$21.77  & $-$22.49  &  10.08\,$\pm$\,0.28 & 0.2\,$\pm$\,0.2    & $-$0.027$_{-0.018}^{+0.010}$ \\
SHiZELS-10  & 1.6\,$\pm$\,0.2        & 10                & 2.3\,$\pm$\,0.2     & 0.13\,$\pm$\,0.04  &  $-$21.15  & $-$21.81  &   9.42\,$\pm$\,0.33 & 0.3\,$\pm$\,0.2    & $-$0.031$_{-0.014}^{+0.016}$ \\
SHiZELS-11  & 1.3\,$\pm$\,0.1        & 8                & 1.3\,$\pm$\,0.4     & 0.6\,$\pm$\,0.1    &  $-$22.13  & $-$24.44  &  11.01\,$\pm$\,0.24 & 0.5\,$\pm$\,0.2    & $-$0.0870$_{-0.006}^{+0.032}$\\
SHiZELS-12  & 0.7\,$\pm$\,0.1        & 5                 & 0.9\,$\pm$\,0.5     & 0.27\,$\pm$\,0.03  &  $-$21.34  & $-$22.44  &  10.59\,$\pm$\,0.30 & 0.3\,$\pm$\,0.2    & ... \\                         
SHiZELS-14  & 1.6\,$\pm$\,0.1        & 27                & 4.6\,$\pm$\,0.4     & 0.60\,$\pm$\,0.05  &  $-$23.55  & $-$24.75  &  10.90\,$\pm$\,0.20 & 0.4\,$\pm$\,0.1    & $-$0.024$_{-0.012}^{+0.012}$ \\
\hline
Median      & 1.1\,$\pm$\,0.1        & 7\,$\pm$\,2          & 2.4\,$\pm$\,0.7     & 0.27\,$\pm$\,0.08  &  $-$21.3\,$\pm$\,0.20 & $-$22.2\,$\pm$\,0.3 & 10.25\,$\pm$\,0.50 & 0.3\,$\pm$\,0.1 & $-$0.026\,$\pm$\,0.006\\
\hline
\label{table:Gal_props}
\end{tabular}
}
\caption{Notes: H$\alpha$ fluxes are in units of
  10$^{-16}$\,erg\,s$^{-1}$\,cm$^{-2}$.  The errors on the fluxes are
  dominated by $\sim$10\% systematic uncertainties in the flux
  calibration from the change in Strehl ratio between observing
  standard stars on-axis and the target galaxies off axis.  r$_{1/2}$
  is the H$\alpha$ half light radius and has been deconvolved for the
  PSF.  Errors on M$_{\rm B}$ \& M$_{\rm K}$ are typically 0.15\,dex
  \citep{Sobral11}.}
\end{center}
\end{table*}

\section{Sample Selection, Observations \& Data Reduction}
\label{sec:obs_dr}

\subsection{HiZELS}

To select the targets for IFU observations, we exploited the high
density of sources in the HiZELS imaging of the COSMOS and UDS fields
\citep{Sobral12b} to select H$\alpha$ emitters which are close to
bright (R\,$<$\,15.0) stars, such that natural guide star adaptive
optics correction could be applied to achieve high spatial resolution.
For this program, we selected fourteen galaxies with H$\alpha$ fluxes
from the HiZELS catalog with fluxes between
0.7--1.6\,$\times$\,10$^{-16}$\,erg\,s$^{-1}$\,cm$^{-2}$ to ensure that
their star-formation properties and dynamics could be mapped in a few
hours.  However, of the fourteen galaxies in our parent sample, only
nine systems yield a H$\alpha$ detection in the data-cubes, which may
(in part) be due to spurious detections in the early HiZELS catalogs.
At the three redshift ranges of our sample, the average H$\alpha$
fluxes of our galaxies correspond to star-formation rates (adopting the
the \citealt{Kennicutt98} calibration with a Chabrier IMF) of
SFR\,=\,4.6\,$\times$\,10$^{-44}$\,L$_{\rm H\alpha}$\,=\,1.7, 6.9 and
19.3\,M$_{\odot}$\,yr$^{-1}$ at $z$\,=\,0.83, 1.47 and 2.23
respectively.

Next we estimate the far-infrared luminosities for the galaxies in our
sample.  We exploit the {\it Herschel}/SPIRE imaging of these fields
which were taken as part of the {\it Herschel} Multi-tiered
Extra-galactic Survey (HerMES).  The SPIRE maps reach 3\,$\sigma$
limits of $\sim$8, 10 and 11\,mJy at 250, 350 and 500$\mu$m
respectively.  None of the SINFONI-HiZELS (SHiZELS) galaxies in our
sample are individually detected (which is unsurprising given the
implied star-formation rates from H$\alpha$), but we can search for an
average star-formation rate via stacking.  

In the following analysis, we restrict ourselves to the $z$\,=\,1.47
galaxies only to mitigate against redshift effects in the stack on the
SED.  The 250, 350 \& 500$\mu$m stacked fluxes for the $z$\,=\,1.47
SHiZELS galaxies of 6.9\,$\pm$\,1.7, 9.0\,$\pm$\,1.9 and
7.0\,$\pm$\,2.4\,mJy at 250, 350 and 500$\mu$m respectively (or
7.5\,$\pm$\,1.6, 9.2\,$\pm$\,2.2 \& 7.4\,$\pm$\,3.0\,mJy at 250, 350 \&
500$\mu$m if we instead stack all nine SHiZELS galaxies in our sample).
A simple modified black-body fit to this photometry (with a dust
emmisivity of $\beta$\,=\,1.5--2.0) at the known redshift yields a
far-infrared luminosity (integrated between 8--1000$\mu$m) of
L$_{FIR}$\,=\,1.8\,$\pm$\,0.6\,$\times$\,10$^{11}$\,L$_{\odot}$
and hence a total star-formation rate of SFR$_{\rm
  FIR}$\,=\,2.7\,$\times$\,10$^{-44}$\,L$_{\rm
  FIR}$\,=\,18\,$\pm$\,8\,M$_{\odot}$\,yr$^{-1}$.

We note that if we instead stack all 317 HiZELS H$\alpha$ emitters from
the $z$\,=\,1.47 parent sample with fluxes
$>$1\,$\times$\,10$^{-16}$\,erg\,s$^{-1}$\,cm$^{-2}$ in the UDS and
COSMOS fields then the 250, 350 \& 500$\mu$m fluxes are
6.3\,$\pm$\,0.24, 7.9\,$\pm$\,0.25 and 5.6\,$\pm$\,0.31\,mJy
respectively, suggesting L$_{\rm
  FIR}$\,=\,1.5\,$\times$\,10$^{11}$\,L$_{\odot}$ which is consistent
with our SHiZELS sample.

Next, for all of the SHiZELS galaxies, we use the broad-band imaging of
the galaxies in order to calculate the rest-frame spectral energy
distribution (SED), reddening and star-formation histories and stellar
masses \citep{Sobral10}.  We exploit deep archival multi-wavelength
imaging from rest-frame UV to mid-infrared bands (see
\citealt{Sobral10} for references) and follow \citet{Sobral11} to
perform a full SED $\chi^2$ fit using the \citet{Bruzual03} and Bruzual
(2007) population synthesis models which include the Thermally
Pulsating Asymptotic Giant Branch (TP-AGB) stellar phase and a Chabrier
IMF (see \citealt{Sobral10} for the range of parameters).  We use
photometry from up to 36 (COSMOS) and 16 (UDS) wide, medium and narrow
bands (spanning {\it GALEX} far-UV and near-UV bands to {\it
  Spitzer}/IRAC; Fig~\ref{fig:SEDs}) and in Table~2 we give the
best-fit stellar mass and E(B$-$V) for each galaxy.  

The average E(B$-$V) for our sample is E(B$-$V)\,=\,0.3\,$\pm$\,0.1
which corresponds to A$_{\rm H\alpha}$\,=\,0.91\,$\pm$\,0.21 (A$_{\rm
  v}$\,=\,1.11\,$\pm$\,0.27).  Applying this correction to the
H$\alpha$ luminosities suggests reddening corrected star-formation
rates SFR$_{\rm H\alpha}$\,=\,16\,$\pm$\,5\,M$_{\odot}$\,yr$^{-1}$,
which is similar to those derived from the far-infrared (SFR$_{\rm
  FIR}$\,=\,18\,$\pm$\,8\,M$_{\odot}$\,yr$^{-1}$).

The rest-frame $B$-band magnitudes (uncorrected for extinction) and
stellar masses are given in Table~2 (the 1$\sigma$ errors are drawn
from the multidimensional $\chi^2$ distribution and are typically
0.15\,dex).  So that more direct comparisons can be made with other
datasets, we also report the rest-frame $B$- and $K$-band magnitudes
for each of the galaxies in our sample, and we note that the average
inferred light-to-mass ratio is L$_{\rm K}$/M\,=\,0.58\,$\pm$\,0.14.

Thus, overall, these galaxies appear to have luminosities consistent
with local LIRGs, with dust corrections of A$_{\rm V}\gsim1$ and
star-formation rates characteristic of ``typical'' star-forming
galaxies at these epochs \citep{Sobral12b}.  Moreover, the SHiZELS
galaxies sample a range of stellar masses from
$\sim$10$^{9-11}$\,M$_{\odot}$ and specific star-formation rates
0.1--10\,Gyr$^{-1}$.  Since one of the main high-redshift comparison
samples we use throughout this paper is the SINFONI Near-Infrared
Galaxy Survey (SINS; \citealt{ForsterSchreiber09}), we note that our
sample has a comparable range of stellar masses, although the SINS
sample tend to have higher H$\alpha$-derived star-formation rates,
(SFR$_{\rm H\alpha}$\,=\,7\,$\pm$\,2\,M$_{\odot}$\,yr$^{-1}$ and
30\,$\pm$\,3\,M$_{\odot}$\,yr$^{-1}$ for SHiZELS and SINS
respectively).

\subsection{SINFONI Observations}

To map the nebular H$\alpha$ emission line properties of the galaxies
in our sample, we used the SINFONI IFU on the ESO VLT
\citep{Eisenhauer03}.  The SINFONI IFU uses an image slicer and mirrors
to reformat a field of 3\,$\times$\,3$''$ at a spatial resolution of
0.1$''$/pixel.  At $z$\,=\,0.84, 1.47 and 2.23 the H$\alpha$ emission
line is redshifted to $\sim$\,1.21, 1.61 and 2.12$\mu$m and into the
$J$, $H$ and $K$-bands respectively.  The spectral resolution in each
band is $\lambda$\,/\,$\Delta\lambda\sim$\,4500 (the sky lines have
4-\AA\ FWHM).  In the remainder of our analysis, emission line widths
are deconvolved for the instrumental resolution.  Since each of the
targets were selected to be close to a bright AO star, we used NGS
correction.  The $R$-band magnitude of the AO stars are
$R$\,=\,12.0--15.0 and they have separations between 8$''$ and 40$''$
from the target galaxies.

To observe the targets we used ABBA chop sequences, nodding 1.5$''$
across the IFU.  We observed each target for between 3.6 and 13.4\,ks
(each individual exposure was 600 seconds) between 2009 September 10
and 2011 April 30 in $\sim$0.6$''$ seeing and photometric conditions.
The median Strehl achieved for our observations is 20\% (Table~1) and
the median encircled energy within 0.1$''$ is 25\% (the approximate
spatial resolution is 0.1$''$ FWHM or 850\,pc at $z$\,=\,1.47 -- the
median redshift of our survey).  Individual exposures were reduced
using the SINFONI {\sc esorex} data reduction pipeline which extracts,
flat-fields, wavelength calibrates and forms the data-cube for each
exposure.  The final data-cube was generated by aligning the individual
data-cubes and then combined these using an average with a 3-$\sigma$
clip to reject cosmic rays.  For flux calibration, standard stars were
observed each night either immediately before or after the science
exposures.  These were reduced in an identical manner to the science
observations.

\begin{figure*}
\centerline{\psfig{file=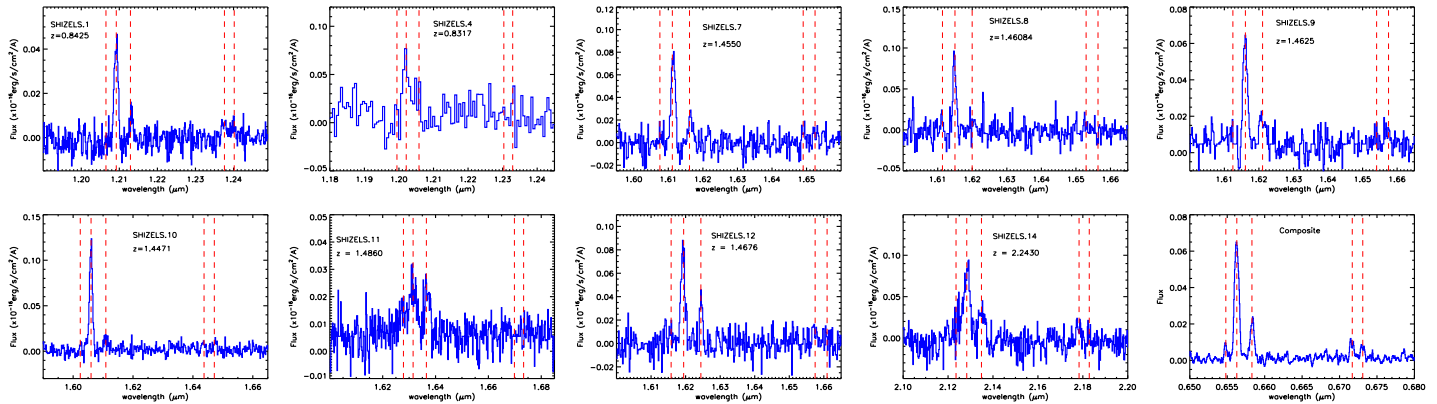,angle=0,width=7.5in}}
\caption{Spatially integrated, one dimensional spectra around the
  redshifted H$\alpha$ emission for each of the galaxies in our sample.
  In each case, the dashed lines identify the H$\alpha$, [N{\sc
      ii}]$\lambda\lambda$\,6583,6548 and [S{\sc
      ii}]$\lambda\lambda$\,6716,6731 emission lines.  In the final
  panel we show the flux weighted composite spectrum.  In all cases, we
  clearly detect [N{\sc ii}] emission and the median [N{\sc
      ii}]\,/\,H$\alpha$ for the sample is 0.3\,$\pm$\,0.1, with a
  range of 0.1\,$<$\,[N{\sc ii}]\,/\,H$\alpha<$\,0.6.  None of the
  galaxies display strong AGN signatures in their near-infrared spectra
  (e.g.\ broad lines or strong [N{\sc ii}]\,/\,H$\alpha$ ratios).}
\label{fig:1dspec}
\end{figure*}

\subsection{Galaxy Dynamics}
\label{sec:dynamics}

All nine galaxies display strong H$\alpha$ emission, with a range of
H$\alpha$ luminosities of L$_{\rm
  H\alpha}\sim$\,10$^{41.4-42.4}$\,erg\,s$^{-1}$.  In
Figure~\ref{fig:1dspec} we show the galaxy-integrated one-dimensional
spectrum from each galaxy.  In all cases we clearly detect the [N{\sc
    ii}]$\lambda$\,6583 emission, the median [N{\sc ii}]\,/\,H$\alpha$ for
the sample is 0.3\,$\pm$\,0.1, with a range of 0.1\,$<$\,[N{\sc
    ii}]\,/\,H$\alpha$$<$\,0.6.  None of the galaxies display strong AGN
signatures in their near-infrared spectra.

To search for fainter lines we also transform each of the spectra to
the rest-frame and coadd them (weighted by flux) and show this stack in
Fig.~\ref{fig:1dspec}.  In this composite, the [N{\sc
    ii}]\,/\,H$\alpha$\,=\,0.29\,$\pm$\,0.02, is consistent with the
average of individual measurements.  We also make a clear detection of
the [S{\sc ii}]$\lambda\lambda$\,6716,6761 doublet and derive a flux
ratio of I$_{6716}$\,/\,I$_{6731}$\,=\,1.36\,$\pm$\,0.08 which
corresponds to an electron density in the range 100--1000\,cm$^{-3}$
\citep{Osterbrock89} and a mass in the ISM of
3--30\,$\times$\,10$^{10}$\,M$_{\odot}$ for a half-light radius of
2\,kpc (Table~2).  It is also useful to note that the [S{\sc
    ii}]\,$\lambda$6716\,/\,H$\alpha$ reflects the ionisation strength
of the ISM and we derive [S{\sc
    ii}]\,$\lambda$6716\,/\,H$\alpha$\,=\,0.13\,$\pm$\,0.04 which
suggests an ionisation parameter
$\log$(U\,/\,cm$^{3}$)\,=\,$-$3.5\,$\pm$\,0.3
\citep{Osterbrock89,CollinsRand01}.  This is comparable to an
ionisation rate of $\log_{10}$(U\,/\,cm$^{3}$)\,$\sim$\,$-$3.9 which
can be inferred from the average star-formation rate volume density of
our galaxies and assuming there are 3\,$\times$\,10$^{60}$ ionising
photons per solar mass of star-formation \citep{Cox83}.

To measure the velocity structure in each galaxy, we fit the H$\alpha$
and [N{\sc ii}]$\lambda\lambda$6548,6583 emission lines pixel-by-pixel.
We use a $\chi^{2}$ minimisation procedure, taking into account the
greater noise at the positions of the sky lines.  We first attempt to
identify a line in each 0.1\,$\times$\,0.1$''$ pixel
($\sim$\,1\,$\times$\,1\,kpc, which corresponds to the approximate
PSF), and if the fit fails to detect the emission line, the region is
increased to 2\,$\times$\,2 pixels.  Using a continuum fit we required
a signal-to-noise $>$5 to detect the emission line in each pixel, and
when this criterion is met we fit the H$\alpha$ and [N{\sc
    ii}]$\lambda\lambda$\,6548,6583 emission allowing the centroid,
intensity and width of the Gaussian profile to vary (the FWHM of the
H$\alpha$ and [N{\sc ii}] lines are coupled in the fit).  Uncertainties
on each parameter of the fit are calculated by perturbing one parameter
at a time, allowing the remaining parameters to find their optimum
values, until $\Delta\chi^2$\,=\,1 is reached.

Even with $\sim$kpc-scale resolution, we note that there is a
contribution to the line widths of each pixel from the large-scale
velocity motions across the PSF which must be corrected for
\citep{Davies11}.  To account for this ''beam smearing'', at each pixel
where the H$\alpha$ emission is detected, we calculate the local
velocity gradient ($\Delta$\,V\,\/\,$\Delta$\,R) and subtract this in
quadrature from the measured velocity dispersion.

In Fig.~\ref{fig:2dmaps} we show the H$\alpha$ intensity, velocity and
line of sight velocity dispersion maps for the galaxies in our sample.
There is clearly a variety of H$\alpha$ morphologies, ranging from
compact (e.g.\ SHiZELS\,11 \& 12) to very extended/clumpy
(e.g.\ SHiZELS\,7, 8, 9, 14).  For three galaxies, we also include
high-resolution broad-band imaging thumbnails from {\it Hubble Space
  Telescope} ACS $V$-band (SHiZELS\,4\&14) or WFC3 $H$-band
(SHiZELS\,11) which shows that the H$\alpha$ is well correlated with
the rest-frame UV/optical morphology.  The distribution and intensity
of star-formation (and the properties of the star-forming clumps) are
discussed in a companion paper (Swinbank et al.\ 2012 ApJ submitted).
Fig.~\ref{fig:2dmaps} also shows that there are strong velocity
gradients in many cases (e.g. SHiZELS\,1, 7, 8, 9, 10, 11, \& 12).  We
will investigate how many of these systems can be classified as
``disks'' in \S3.

\begin{figure*}
  \centerline{\psfig{file=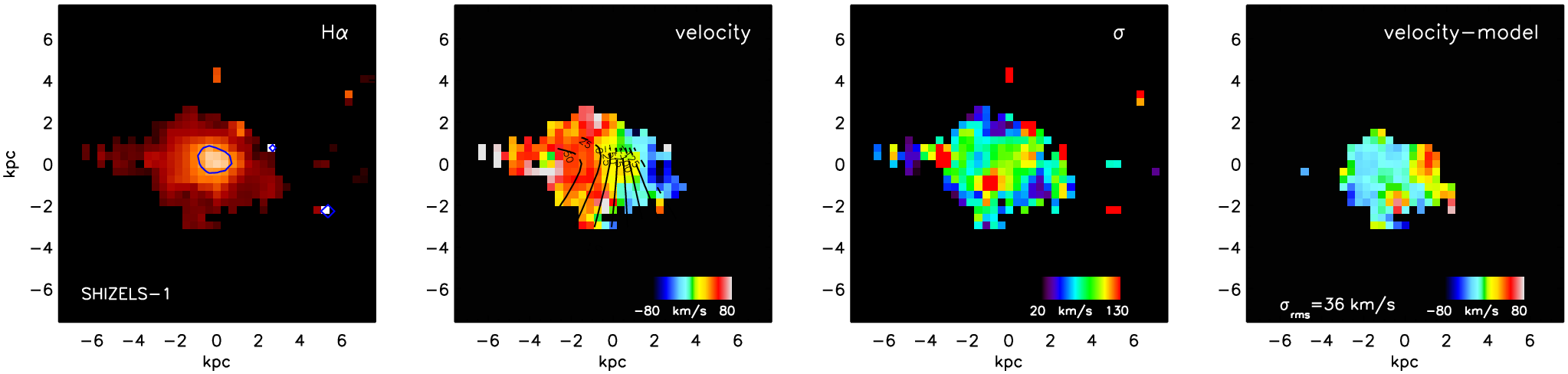,angle=0,width=6.5in}}
  \centerline{\psfig{file=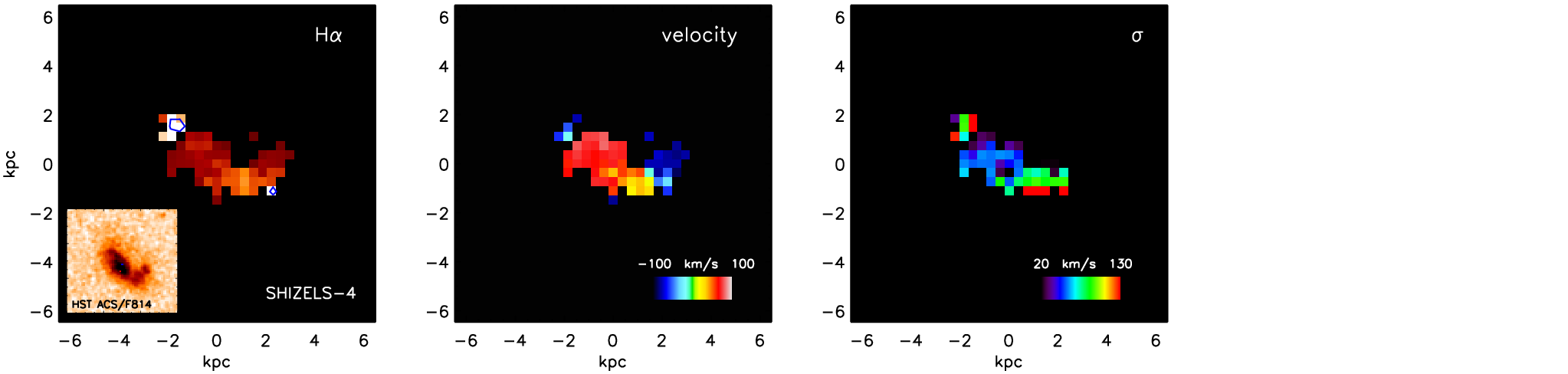,angle=0,width=6.5in}}
  \centerline{\psfig{file=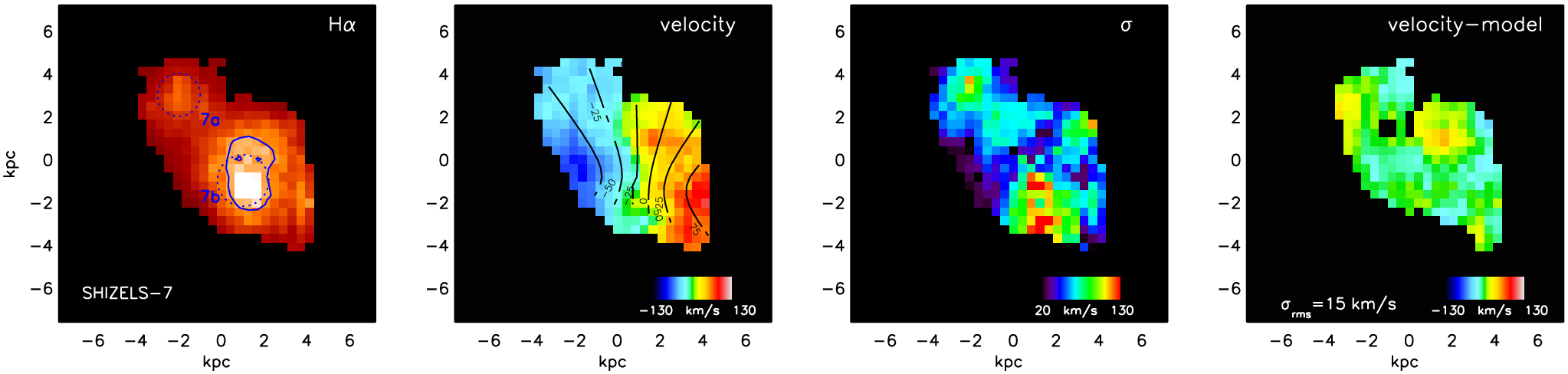,angle=0,width=6.5in}}
  \centerline{\psfig{file=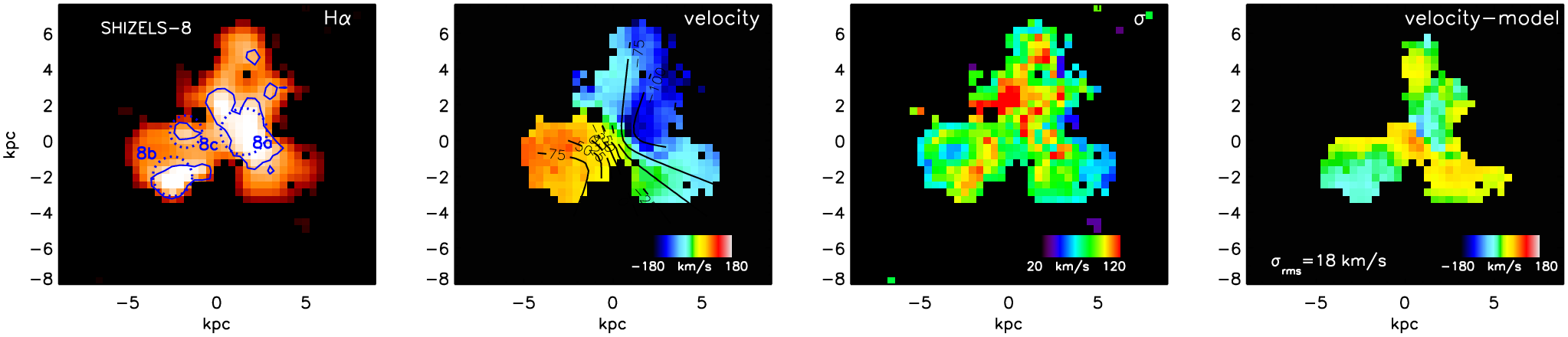,angle=0,width=6.5in}}
  \centerline{\psfig{file=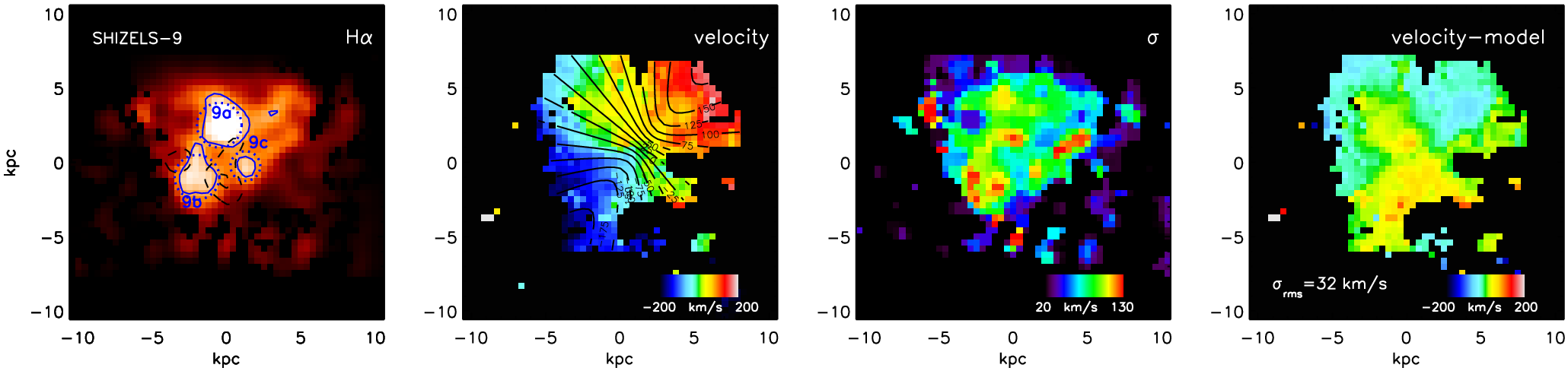,angle=0,width=6.5in}}
\caption{H$\alpha$ and kinematic maps of the SHiZELS galaxies.  For
  each galaxy, the left hand panel shows the H$\alpha$ emission line
  flux.  The contours denote a H$\alpha$-derived star-formation surface
  density of $\Sigma_{\rm
    SF}$\,=\,0.1\,M$_{\odot}$\,yr$^{-1}$\,kpc$^{-2}$.  The central two
  panels show the velocity field and line-of-sight velocity dispersion
  profile ($\sigma$) respectively.  The right hand panel shows the
  residual velocity field after subtracting the best-fit kinematic
  model.  The r.m.s. of these residuals is given in each panel (for
  SHiZELS 4 \& 12 there are too few resolution elements across the
  source to meaningfully attempt to fit disk models).}
\label{fig:2dmaps}
\end{figure*}

\addtocounter{figure}{-1}
\begin{figure*}
  \centerline{\psfig{file=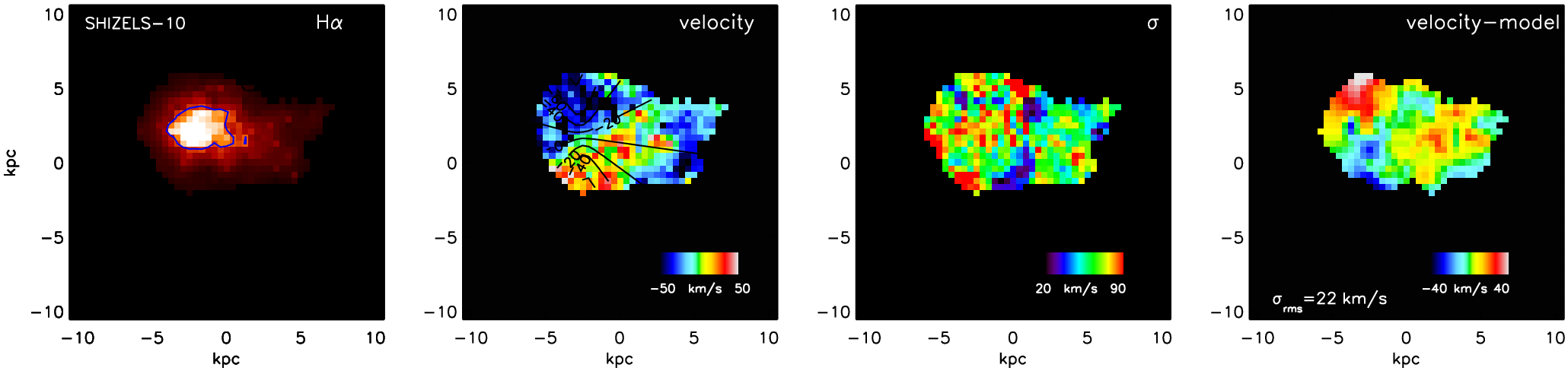,angle=0,width=6.5in}}
  \centerline{\psfig{file=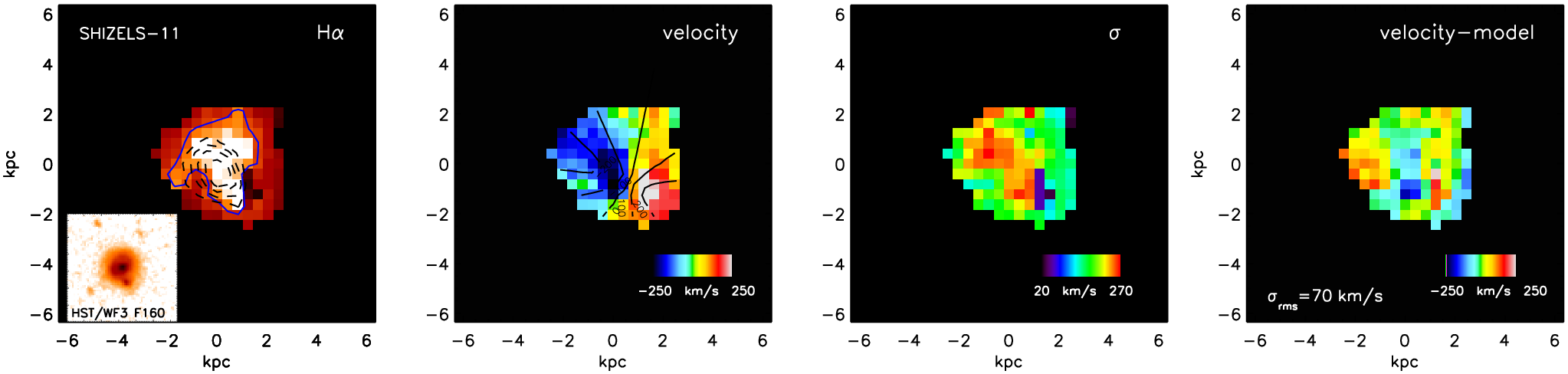,angle=0,width=6.5in}}
  \centerline{\psfig{file=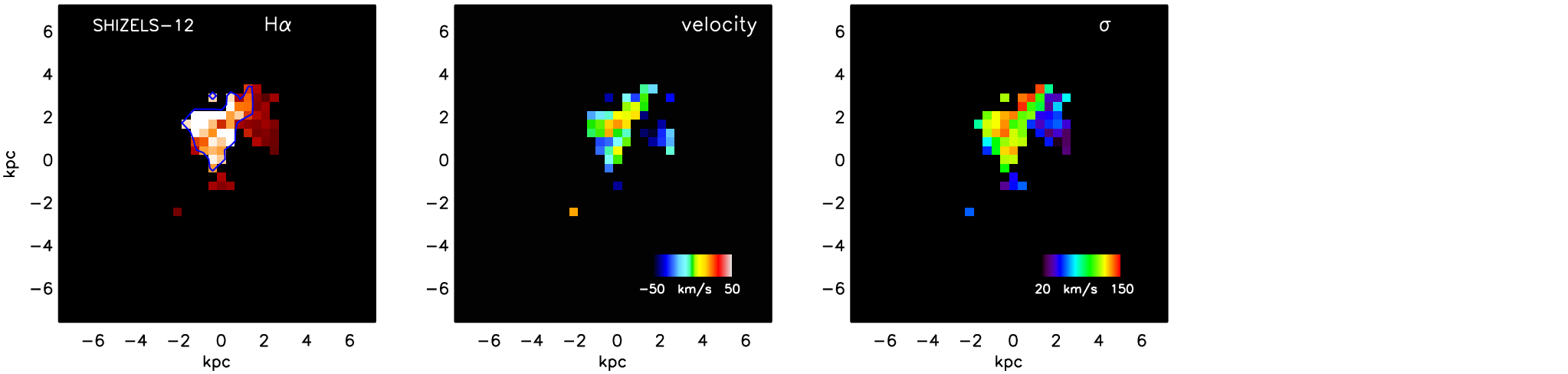,angle=0,width=6.5in}}
  \centerline{\psfig{file=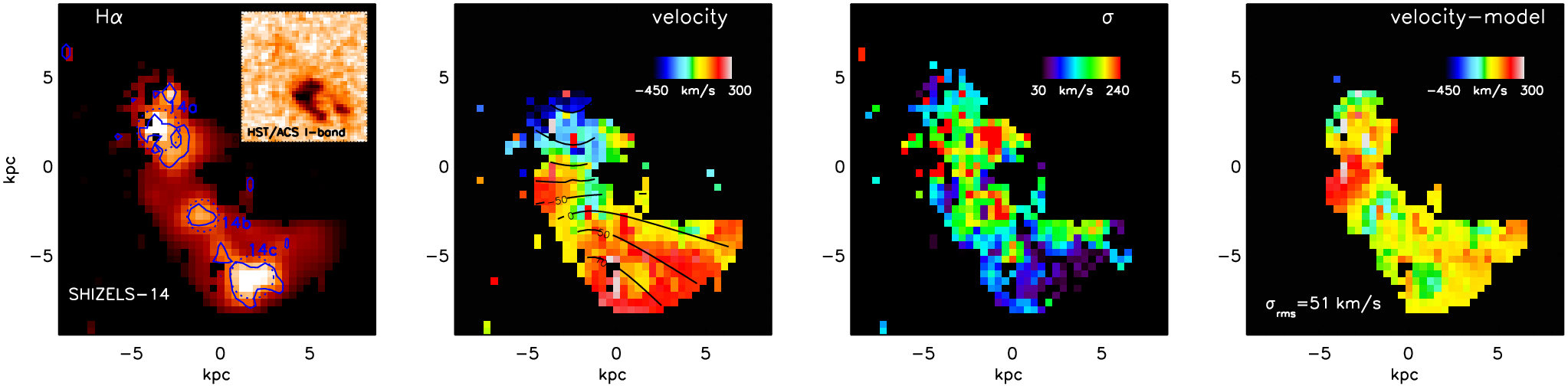,angle=0,width=6.5in}}
\caption{continued...}
\label{fig:2dmapsB}
\end{figure*}

\begin{figure}
  \centerline{\psfig{file=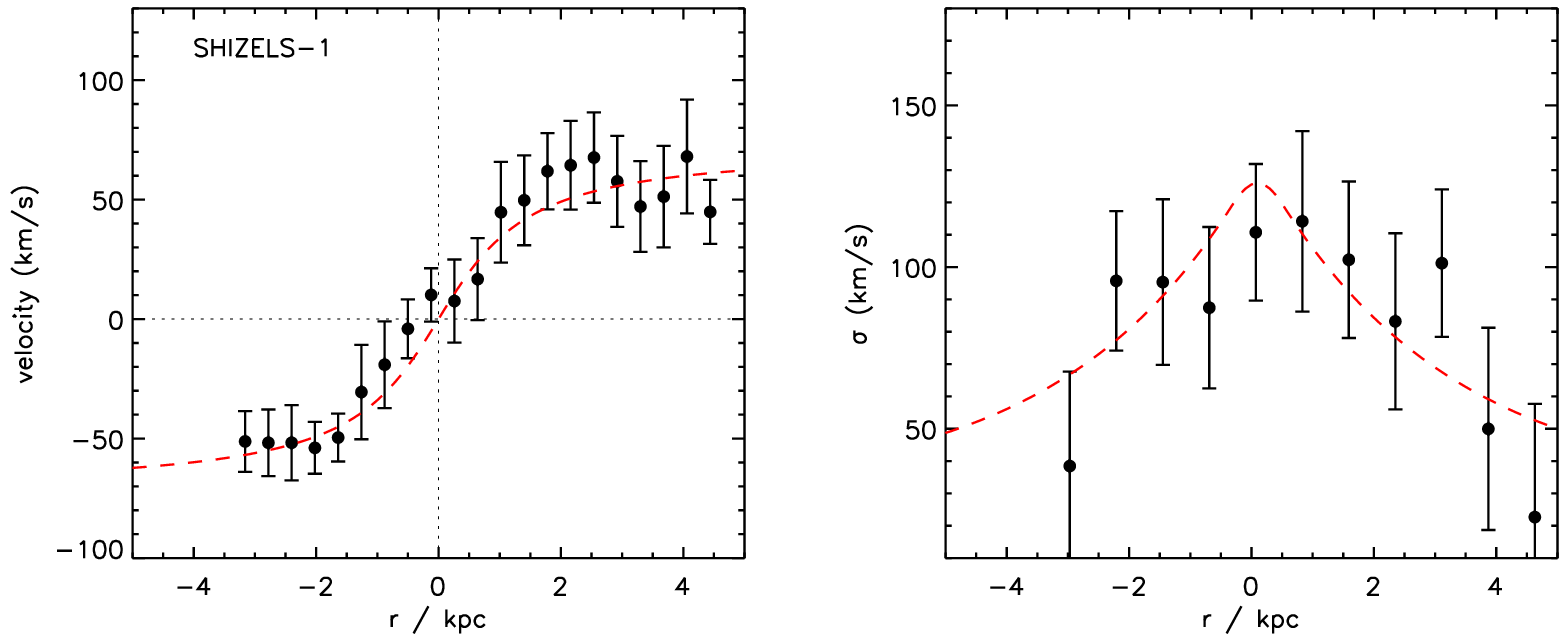,angle=0,width=2.9in}}
  \centerline{\psfig{file=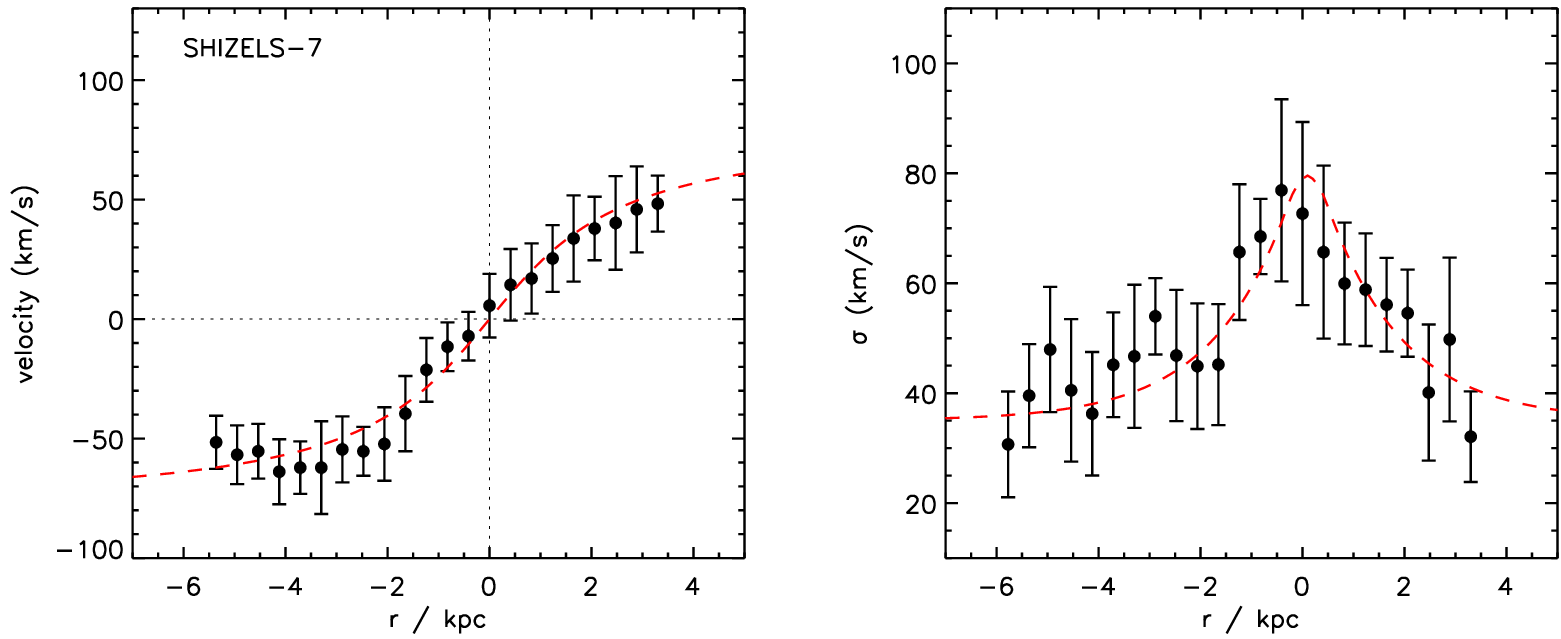,angle=0,width=2.9in}}
  \centerline{\psfig{file=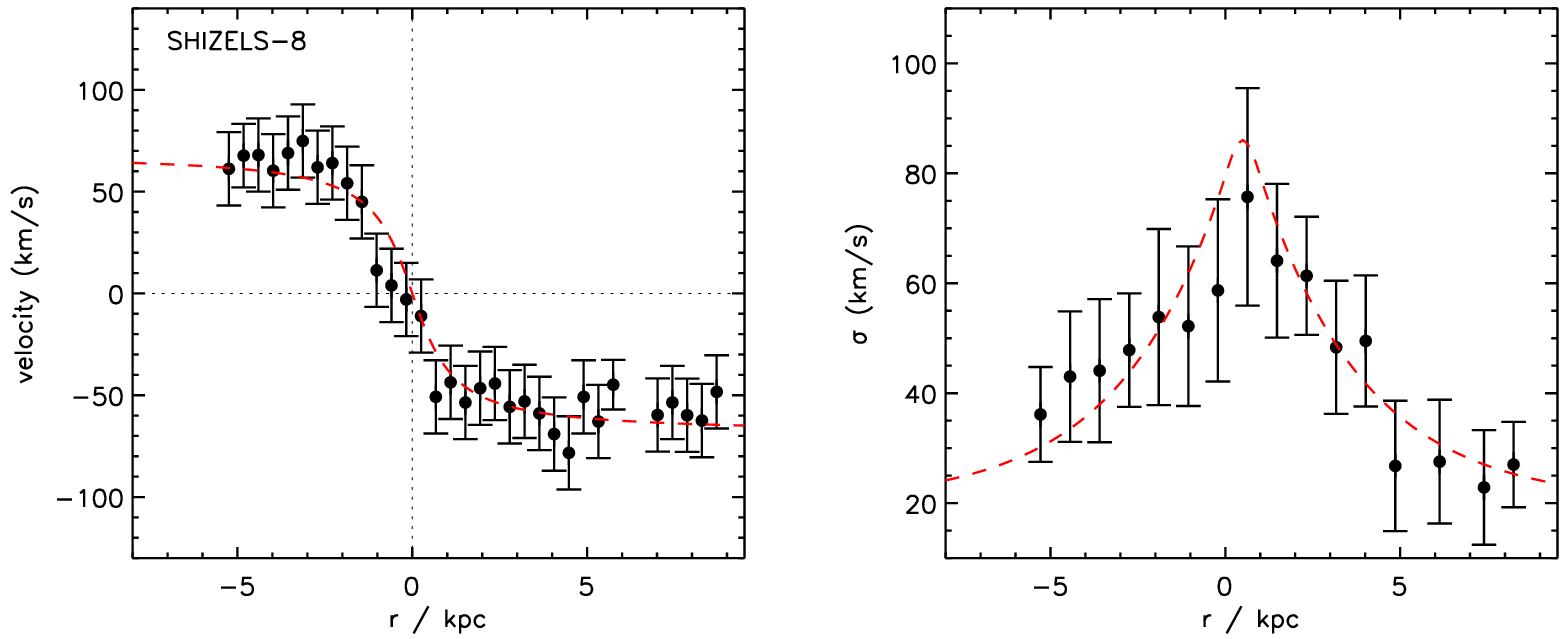,angle=0,width=2.9in}}
  \centerline{\psfig{file=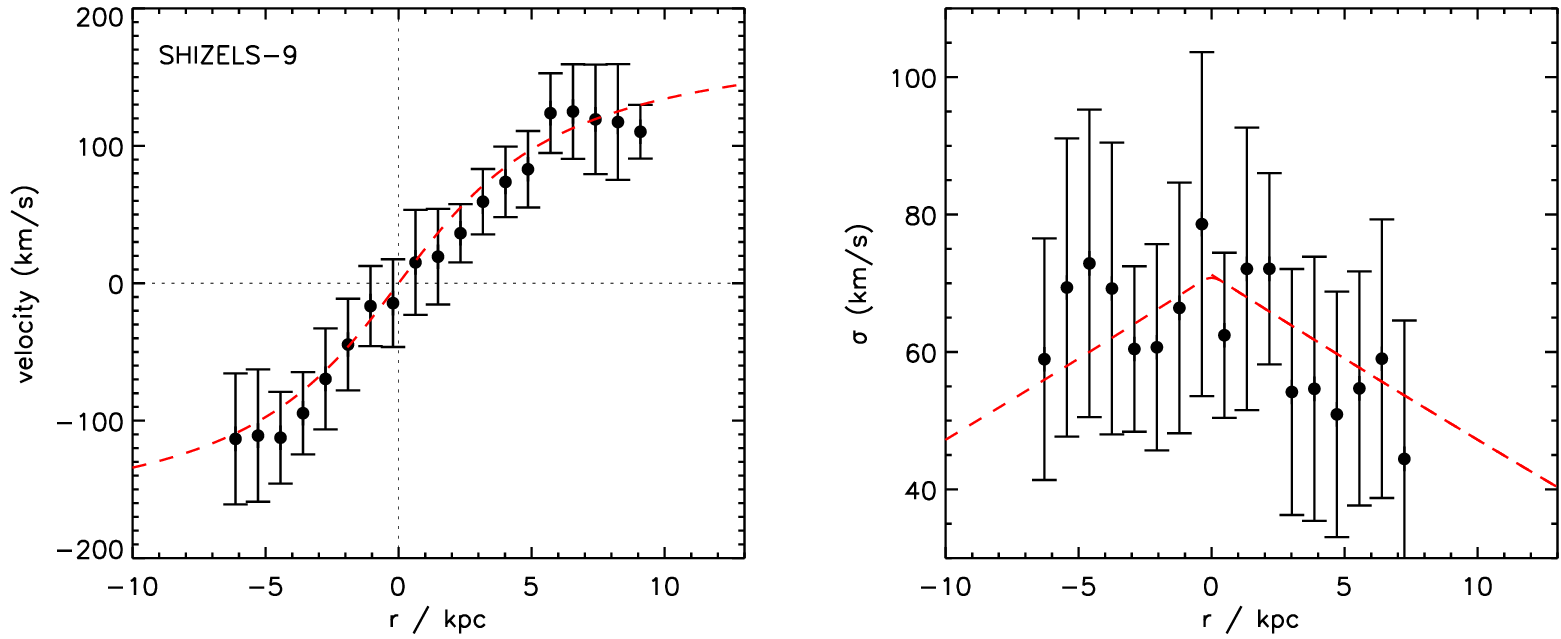,angle=0,width=2.9in}}
  \centerline{\psfig{file=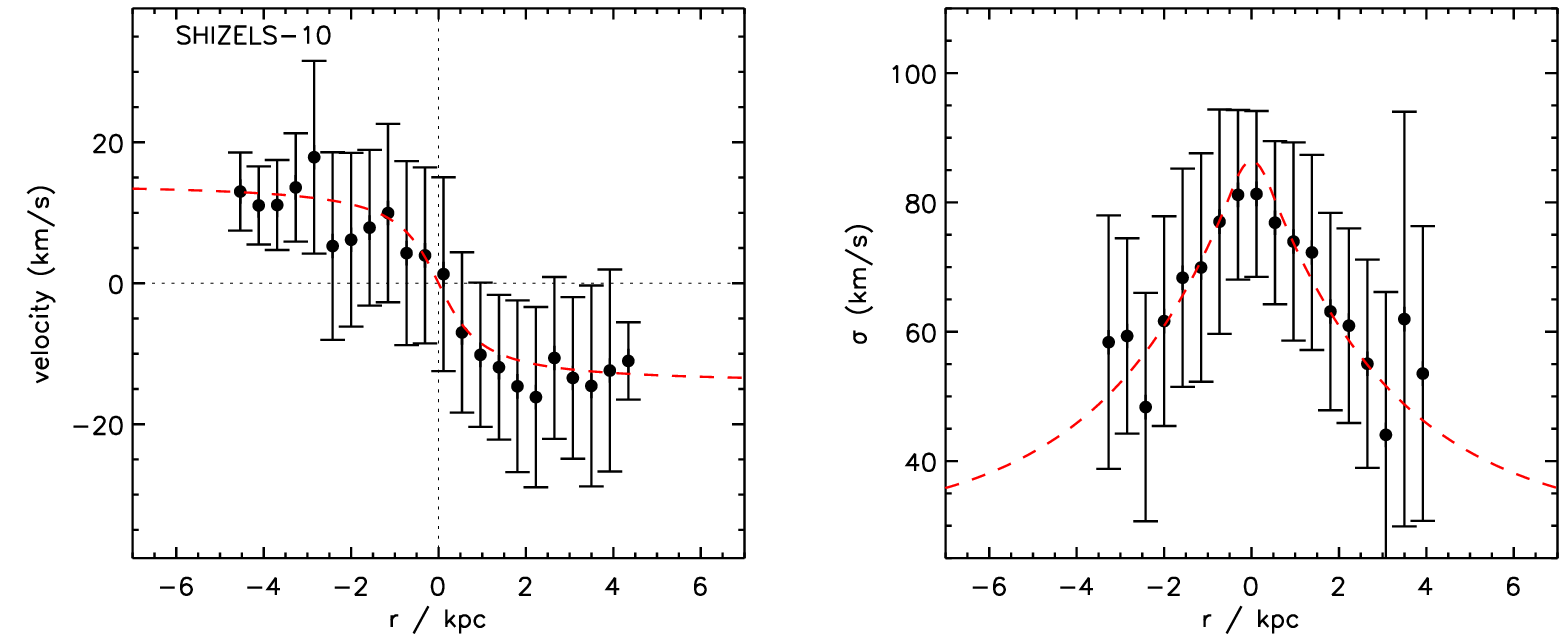,angle=0,width=2.9in}}
  \centerline{\psfig{file=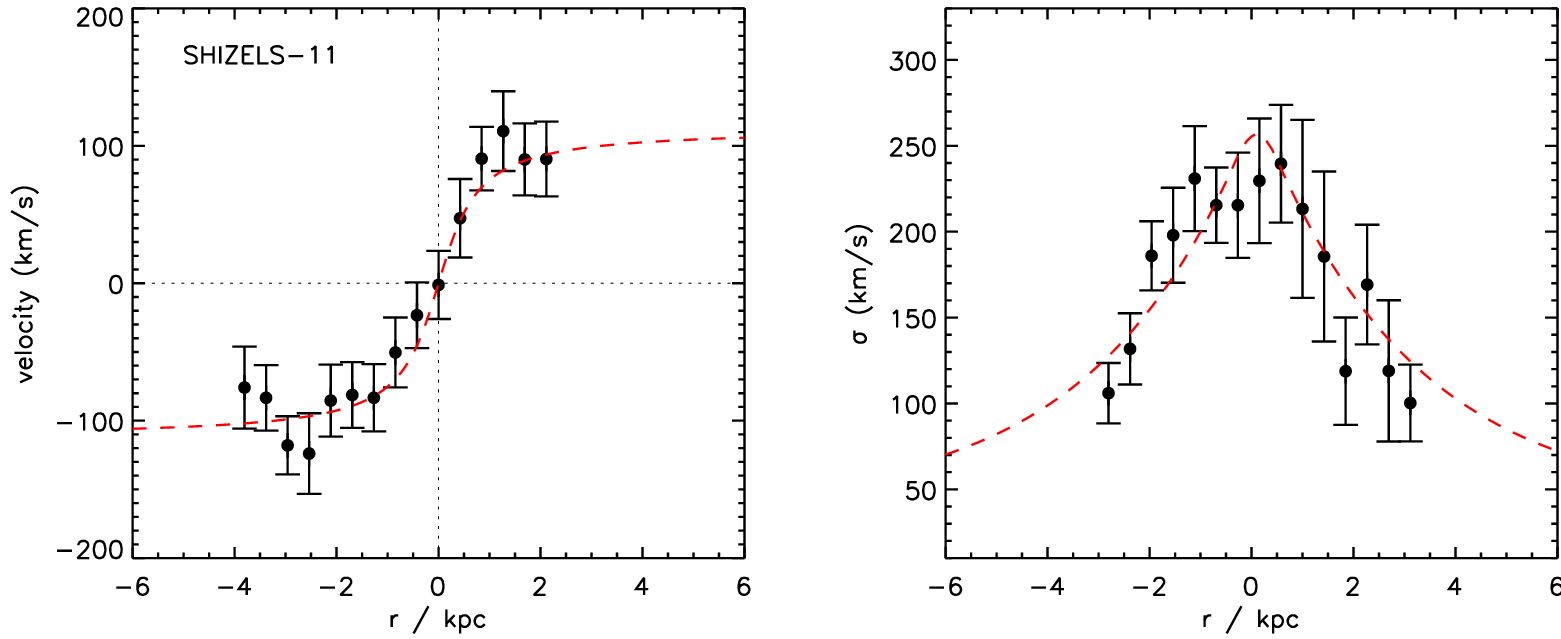,angle=0,width=2.9in}}
  \centerline{\psfig{file=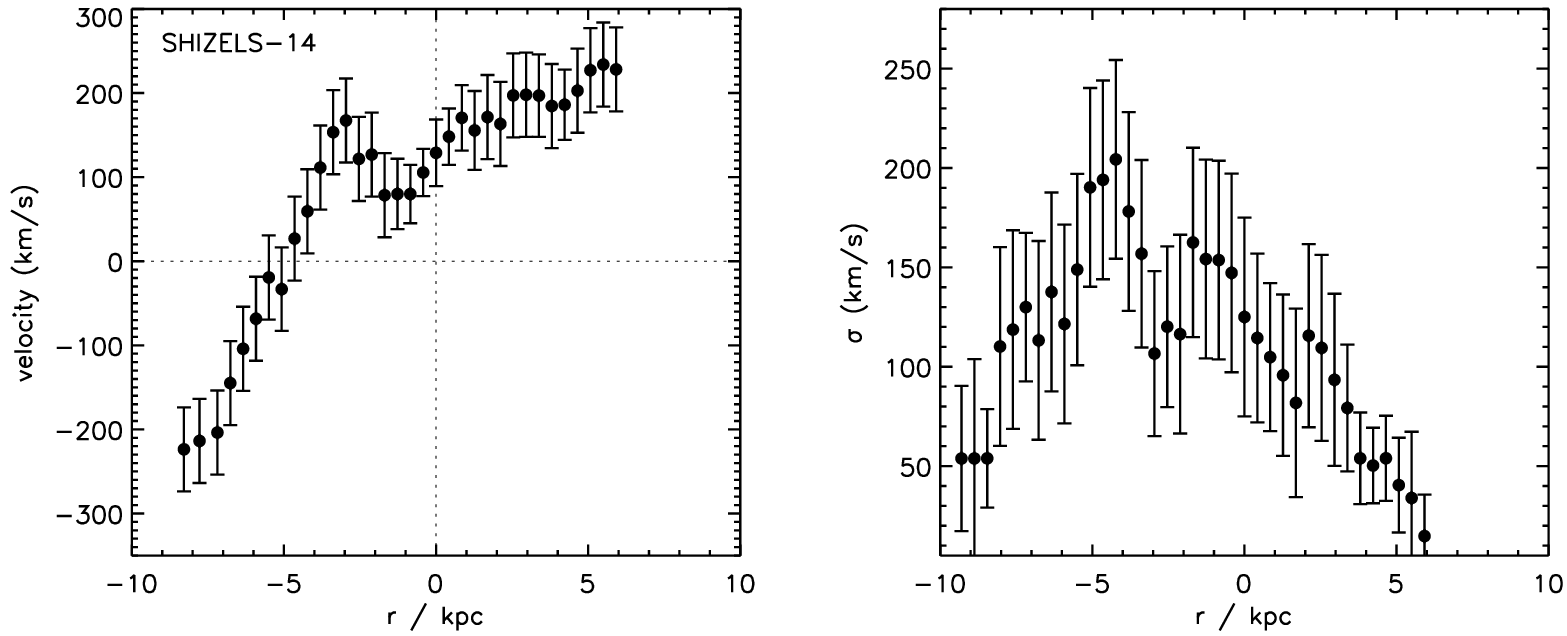,angle=0,width=2.9in}}
\caption{{\it Left:} One dimensional rotation curves for the seven
  galaxies in our sample which show strong velocity gradients in their
  H$\alpha$ dynamics.  For the six galaxies whose dynamics resemble
  rotating systems, we overlay the best-fit one dimensional rotation
  curve {\it Right:} One dimensional velocity dispersion profiles with
  the best-fit line of sight profile overlaid for the rotating systems.
  The observed rotation curve and line-of-sight velocity dispersion are
  extracted from the two-dimensional velocity fields using a 1\,kpc
  width slit across the major kinematic axis of each galaxy.}
\label{fig:1dprofiles}
\end{figure}

\begin{table*}
\begin{center}
{\small
{\centerline{\sc Table 3: Dynamical Properties}}
\begin{tabular}{lccccccccc}
\hline
\noalign{\smallskip}
ID          & v$_{\rm asym}$ & $\sigma$  &V$_{2.2}$   & Inc & K$_{\rm V}$     & K$_{\sigma}$    & K$_{\rm Tot}$    &$\chi^2_{\nu}$ & Class     \\
            & (km/s)       & (km/s)     &(km/s)     & (deg) &              &                &                 &    &         \\
\hline
SHiZELS-1   & 112$\pm$11 & 98$\pm$15  & 86$\pm$18  & 40   & 0.12$\pm$0.17  & 0.17$\pm$0.81  &  0.20$\pm$1.03  &   4.8 & D   \\ 
SHiZELS-4   & ...        & 77$\pm$20   & ...        & ...  &     ...        &      ...       &        ...      &   ... & C   \\ 
SHiZELS-7   & 145$\pm$10 & 75$\pm$11   & 112$\pm$15 & 33   & 0.10$\pm$0.08  & 0.35$\pm$0.18  &  0.36$\pm$0.34  &   3.1 & D   \\
SHiZELS-8   & 160$\pm$12 & 69$\pm$10   & 144$\pm$17 & 25   & 0.20$\pm$0.15   & 0.30$\pm$0.25 &  0.36$\pm$0.40  &   3.0 & D   \\
SHiZELS-9   & 190$\pm$20 & 62$\pm$11   & 155$\pm$24 & 54   & 0.12$\pm$0.05  & 0.48$\pm$0.20  &  0.49$\pm$0.30  &   5.7 & D   \\
SHiZELS-10  & 30$\pm$12  & 64$\pm$8    & 26$\pm$12  & 30   & 0.13$\pm$0.05  & 0.83$\pm$0.14  &  0.84$\pm$0.35  &   25.5 & M   \\
SHiZELS-11  & 224$\pm$15 & 190$\pm$18  & 208$\pm$18 & 25   & 0.16$\pm$0.05  & 0.34$\pm$0.10  &  0.37$\pm$0.16  &   5.34 & D   \\
SHiZELS-12  & ...        & 115$\pm$10  & ...         & ...  &  ...           &     ...        &        ...      &  ... & C   \\
SHiZELS-14  & ...        & 131$\pm$17  & 190$\pm$25  & 58   & 0.15$\pm$0.09  & 0.38$\pm$0.23  &  0.40$\pm$0.34  & 9.1 & M   \\
\hline       
Median      & 147$\pm$31  & 75$\pm$19  & 133$\pm$30  & ...  & 0.13$\pm$0.01  & 0.37$\pm$0.07  & 0.38$\pm$0.04   & 5.3$\pm$3.0 & ...  \\
\hline
\label{table:Gal_props}
\end{tabular}
}
\caption{Notes: v$_{\rm asym}$ and V$_{2.2}$ are inclination corrected.
  $\sigma$ is the avergae velocity dispersion across the galaxy image
  (and the velocity dispersion map has been corrected for "beam
  smearing" effects due to the PSF, see \S~ref{sec:dynamics}.  The
  classes in the final column denote Disk (D), Merger (M) and Compact
  (C) (see \S3)}
\end{center}
\end{table*}

\section{Analysis, Results \& Discussion}
\label{sec:analysis}

All nine galaxies display velocity gradients in their H$\alpha$
kinematics, with peak-to-peak velocity gradients ranging from
40--400\,km\,s$^{-1}$ and a ratio of peak-to-peak difference ($v_{\rm
  max}$), to line of sight velocity dispersion ($\sigma$) of $v_{\rm
  max}$\,/\,$\sigma$\,=\,0.3--2.9.  Although a ratio of $v_{\rm
  max}$\,/\,$\sigma$\,=\,0.4 has traditionally been used to crudely
differentiate rotating systems from mergers \citep{ForsterSchreiber09},
more detailed kinematic modelling is essential to reliably
differentiate rotating systems from mergers \citep{Shapiro08}.  We
therefore begin by modelling each velocity field with a rotating disk
in order to estimate the disk inclination and true rotational velocity
(we exclude SHiZELS\,4 \& 12 from this analysis as their velocity
fields are only marginally resolved by our data).

Although there have been several functional forms to describe rotation
curves (such as the ``multi-parameter fit''; e.g. \citealt{Courteau97}
or the ``universal rotation curve; \citealt{Persic96}), the simplest
model which provides a good fit to most rotation curves on $\sim$\,kpc
scales is the {\rm arctan} fit of the form
$v(r)$\,=\,$\frac{2}{\pi}$\,$v_{\rm asym}$\,${\rm arctan}(r/r_t)$,
where $v_{\rm asym}$ is the asymptotic rotational velocity and r$_{\rm
  t}$ is the effective radius at which the rotation curve turns over
\citep{Courteau97}.  To fit the two-dimensional velocity fields of the
galaxies in our sample, we construct a two-dimensional kinematic model
with six free parameters (v$_{\rm asym}$, r$_{\rm t}$, [x/y] centre,
position angle (PA) and disk inclination) and use a genetic algorithm
\citep{Charbonneau95} to find the best model.  During the fitting, we
limit the parameters ranges to those allowed by the data (e.g.\ the
[x/y] center and turn over radius, r$_{\rm t}$, must be within the
SINFONI field of view; the maximum circular velocity must be
$<$\,2\,$\times$ the maximum velocity seen in the data;
PA\,=\,0--180$^{\circ}$, and $i$\,=\,0--90).  We then generate 10$^{5}$
models within random initial parameters between these ranges and find
the lowest $\chi^2_{\nu}$.  This process is repeated (up to 100 times),
but in each new generation we remove (up to) 10\% of the worst fits and
contract the remaining parameter space search by up to 5\% centered on
the best-fit.  We examine whether the fit has converged by examining
whether all of the models in a generation are within
$\Delta\chi^2$\,=\,1 of the best-fit model.  The errors on the final
parameters (Table~3) reflect the range of acceptable models from {\it
  all} of the models attempted.

The best-fit kinematic maps and velocity residuals are shown in
Fig.~\ref{fig:2dmaps}, whilst the best-fit inclinations and disk
rotation speeds are given in Table~2, along with the $\chi^2_{\nu}$.
All galaxies show small-scale deviations from the best-fit model, as
indicated by the typical r.m.s,
$<$\,data\,$-$\,model\,$>$\,=\,30\,$\pm$\,10\,km\,s$^{-1}$, with a
range from $<$\,data\,$-$\,model\,$>$\,=\,15--70\,km\,s$^{-1}$.  These
offsets could be caused by the effects of gravitational instability, or
simply due to the unrelexed dynamical state indicated by the high
velocity dispersions ($\sigma$\,=\,96\,$\pm$\,24\,km\,s$^{-1}$).  The
goodness of fit and small scale deviations from the best-fit models are
similar to those of 18 galaxies in the SINS survey which show the most
prominent rotational support (r.m.s of 10--80\,km\,s$^{-1}$ and
$\chi^2_{\nu}$\,=\,0.2--20; \citealt{Cresci09}), as well as to the
high-resolution studies of gravitationally lensed galaxies from
\citet{Jones10} and \citet{Stark08} (which have an average
$\chi^2_{\nu}$\,=\,3.5\,$\pm$\,0.8).  We therefore conclude that the
disk model provides an adequate fit to the majority of these galaxies
and that the velocity fields of most galaxies are consistent with the
kinematics of turbulent rotating disks.

We use the best-fit dynamical model to identify the major kinematic
axis, and extract the one dimensional rotation curve and velocity
dispersion profiles, and show these in Fig.~\ref{fig:1dprofiles}.  We
define the zero point in the velocity using the dynamical center of the
galaxy, whilst the error bars for the velocities are derived from the
formal $1\sigma$ uncertainty in the velocity arising from the Gaussian
profile fits to the H$\alpha$ emission.  We note that in the
one-dimensional rotation curves in Fig.~\ref{fig:1dprofiles} alternate
points show independent data.  We also note that the data have not been
folded about the zero velocity, so that the degree of symmetry can be
assessed.

Whilst this modelling provides a useful test of whether the dynamics
can be described by disk-like rotation, a better criterion for
distinguishing between motion from disturbed kinematics is the
``kinemetry'' which measures the asymmetry of the velocity field and
line-of-sight velocity dispersion maps for each galaxy.  This technique
has been well calibrated and tested at low redshift
\citep[e.g.\ ][]{Krajnovic07}, whilst at high redshift it has been used
to determine the strength of deviations of the observed velocity and
dispersion maps from an ideal rotating disk \citep{Shapiro08}.  In this
modelling, the velocity and velocity dispersion maps are described by a
series of concentric ellipses of increasing semi-major axis length, as
defined by the system center, position angle and inclination.  Along
each ellipse, the moment map as a function of angle is extracted and
decomposed into its Fourier series which have coefficients $k_n$ at
each radii (see \citealt{Krajnovic07} for more details).

In a noiseless, idealised disk, the coefficients of the Fourier
expansion would be zero.  Any deviations from the ideal case that occur
in a disturbed disk (e.g. warps, spiral structures, streaming motions
or mergers in the extreme cases) induce strong variations in the
higher-order Fourier coefficients.  To ensure that reliable
measurements of the kinemetry parameters can be made, we first
construct a grid of 2000 model disks and mergers with noise and spatial
resolution appropriate for our SINFONI observations.  For the model
disks, we allow disk scale lengths and rotation curve turn over lengths
(r$_{\rm t}$) from 0.5--3\,kpc, rotation speeds and peak velocity
dispersions from 10--300\,km\,s$^{-1}$ and random inclinations and sky
position angles.  The model mergers typically comprise 2--3 of these
galaxy-scale components separated by 1--20\,kpc in projection.  The
dynamics of each of these components randomly ranges from dispersion
dominated to rotating systems.

We then use the kinemetry code from \citet{Krajnovic07} and measure the
velocity and velocity dispersion asymmetry for all of these models.
Following \citet{Shapiro08}, we define the velocity asymmetry (K$_{\rm
  V}$) as the average of the $k_n$ coefficients with $n$\,=\,2--5,
normalised to the first Cosine term in the Fourier series (which
represents circular motion); and the velocity dispersion asymmetry
(K$_{\sigma}$) as the average of the first five coefficients
($n$\,=\,1--5) also normalised to the first Cosine term.  For an ideal
disk, K$_v$ and K$_{\sigma}$ will be zero.  In a merging system, strong
deviations from the idealised case causes large K$_v$ and K$_{\sigma}$
values, which can reach K$_{\rm v}\sim$\,K$_{\sigma}\sim$\,10 for very
disturbed systems.  The total asymmetry, K$_{\rm Tot}$ is $K_{\rm
  Tot}^2$\,=\,K$_{\rm V}^2$\,+\,K$_{\sigma}^2$.  For our mock sample of
model disks, we recover K$_{\rm Tot,disk}$\,=\,0.30\,$\pm$\,0.03
compared to K$_{\rm Tot,merger}$\,=\,2\,$\pm$\,1 for the mergers.

For the SHiZELS galaxies in our sample, we measure the velocity and
velocity dispersion asymmetry, (SHiZELS\,4 \& 12 have too few
independent spatial resolution elements across the galaxy so we omit
these from the kinemetry analysis).  \citet{Krajnovic07} show that an
incorrect choice of centre induces artificial power in the derived
kinemetry coefficients.  We therefore allow the dynamical center to
vary over the range allowed by the family of best-fit two dimensional
models and measure the kinemetry in each case.  We also perturb the
velocity and dispersion maps by the errors on each pixel and re-measure
the asymmetry, reporting the velocity and dispersion asymmetries,
(K$_{\rm V}$ and K$_{\sigma}$ respectively) along with their errors in
Table~3.  We use the total asymmetry, K$_{\rm Tot}$ to differentiate
disks (K$_{\rm Tot}<$\,0.5) from mergers K$_{\rm Tot}>$\,0.5).  For the
galaxies in our sample, five have are classified as disks (D), whilst
two more indicate mergers (M), and the final two are compact (C).

As expected, the two galaxies that are classified as mergers from the
kinemetry (SHiZELS\,10 \& 14) have the highest $\chi^2_{\nu}$
($\chi^2_{\nu}>$\,9) from the dynamical modelling.  Both of these
systems are noteworthy; the one-dimensional rotation curve of
SHiZELS\,10 (extracted from the major kinematic axis) appears regular
and suggests rotation, but the two-dimensional velocity field reveals
extended ($\sim$\,5\,kpc) emission to the east, which may be due to a
companion or tidal material.  The dynamics of the other galaxy
classified as a merger, SHiZELS\,14, are clearly more complex.  The
H$\alpha$ is extended across $\sim$\,14kpc with a peak-to-peak velocity
gradient of $\sim$480\,$\pm$\,40\,km\,s$^{-1}$.  Qualitatively, the one- and
two-dimensional velocity field of SHiZELS\,14 are consistent with an
early stage prograde encounter, with each component in the merger
displaying approximately equal peak-to-peak velocity gradients.  The
northern most component is also more highly obscured than the southern
component (as evident from the \emph{HST} $I$-band imaging;
Fig.~\ref{fig:2dmaps}).

Overall, the fraction of rotating systems within our sample,
$\sim$\,55\%, is consistent with that found from other H$\alpha$ IFU
surveys of high-redshift star-forming galaxies
\citep[e.g.\ ][]{ForsterSchreiber09,Jones10,Wisnioski11}.


\subsection{The Tully-Fisher Relation}
\label{sec:TF}

Since our two dimensional dynamical data resolve the turn over in the
rotation curves (as evident from Fig.~\ref{fig:1dprofiles}), we can use
our results to investigate how the disk scaling relations for the
galaxies in our sample compare to galaxy disks at $z$\,=\,0.  The
relation between the rest-frame $B$-band luminosity (M$_{B}$) and
rotational velocity and that between the total stellar-mass
(M$_{\star}$) and rotational velocity define the Tully-Fisher relations
\citep{TullyFisher}.  The first of these relations has a strong
contribution from the short-term star-formation activity whilst the
second provides a better proxy for the integrated star-formation
history.  Indeed the latter relationship may reflect how
rotationally-supported galaxies formed, perhaps suggesting the presence
of self-regulating processes for star-formation in galactic disks.  The
slope, intercept and scatter of the Tully-Fisher relations and their
evolution are therefore key parameters that any successful galaxy
formation model must reproduce
\citep[e.g.\ ][]{Cole89,Cole94,Kauffmann93,Eisenstein96,Somerville99,Steinmetz99,Portinari07,Dutton11}.

Since we wish to compare our dynamical measurements with those of other
high-redshift galaxies where similar data exist, it is important to
measure the rotational velocity in a consistent manner.  Whilst the
{\it arctan} function is useful for describing the overall shape of the
rotation curve (in both one and two-dimensional data), it provides an
extrapolation of the rotational velocity, which may not be warranted
especially if the rotation curve does not flatten in the data.
Equally, since any measurement of the observed peak-to-peak rotational
velocity (V$_{\rm max}$) is sensitive to the radius at which it has
been possible to detect the dynamics, this measurement varies from both
object-to-object depending on both the extent of the galaxy and
signal-to-noise of the observations and can therefore give biased
measurements when comparing samples.

A more robust measurement is that of the velocity at 2.2 times the disk
scale length (V$_{2.2}$).  This has a better physical basis since it
measures the peak rotational velocity for a pure exponential disk
\citep{Freeman70,Courteau97b} and provides the closest match the radio
(21\,cm) line widths for local galaxies \citep{Courteau97}.  Of course,
the light profile of high-redshift galaxies are unlikely to follow pure
exponential disk profiles, but it provides a reasonable estimate of
where the rotation curve flattens, and as we will see below, is a good
approximation of both V$_{\rm max}$ and V$_{\rm asym}$ in our data.  To
measure V$_{2.2}$ in our data, we therefore measure the H$\alpha$
intensity as a function of radius in our galaxies (centered at the
dynamical center and using the inclination and position angle set by
the best fit two-dimensional kinematic model) and infer the disk scale
length (which ranges from 0.7--2.7\,kpc and has a median of
1.8\,$\pm$\,0.7\,kpc).  We then measure the rotational velocity at 2.2
disk scale lengths and report these in Table~3.  The errors on these
quantities are derived from both the errors on the observed velocity at
this radius and the formal errors on the two-dimensional model velocity
fields.  For reference, we note that owing to the high resolution and
signal-to-noise of our data, we have been able to resolve the
flattening in the rotation curves, the ratio of the maximum
(observed) peak-to-peak velocity (V$_{\rm max}$) to V$_{\rm 2.2}$ in
our sample is V$_{max}$\,/\,V$_{2.2}$\,=\,0.93\,$\pm$\,0.05 whilst the
ratio of asymptotic velocity (V$_{\rm asym}$) to V$_{2.2}$ is
V$_{asym}$\,/\,V$_{2.2}$\,=\,1.25\,$\pm$\,0.08.

\subsubsection{B-Band Tully-Fisher Relation}
\label{sec:BTF}

\begin{figure*}
\centerline{\psfig{file=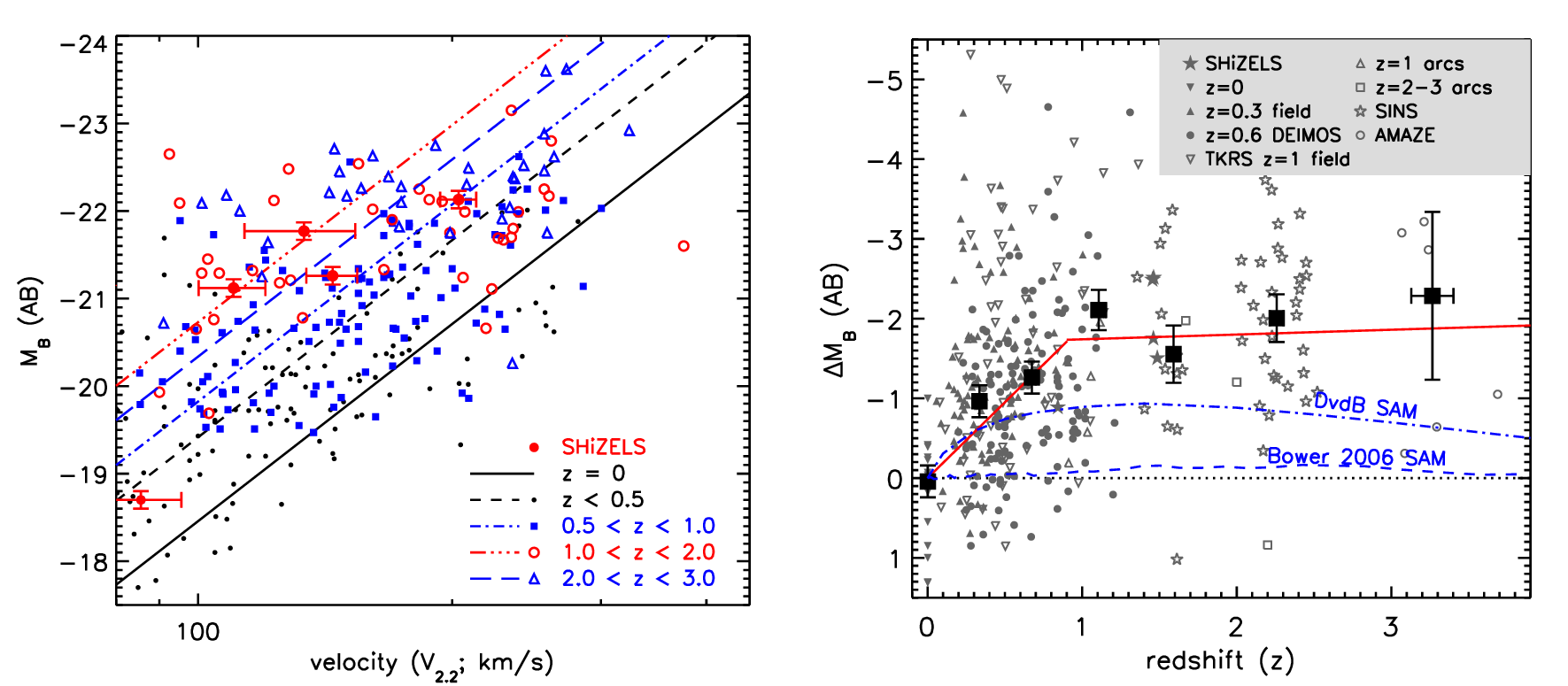,angle=90,width=7in}}
\caption{{\it Left:} The evolution of the rest-frame $B$-band
  Tully-Fisher relation.  We baseline our results against the $z$\,=\,0
  measurements from \citet{Pierce92}.  For the high-redshift samples,
  we combine the observations from this work with a number of other
  surveys.  At $z\sim$\,0.1--1 we include the field surveys from
  \citet{Bamford05} ($z$\,=\,0.3 field), \citet{Weiner06} (TKRS
  $z$\,=\,1 field) and \citet{Miller11} ($z$\,=\,0.6 DEIMOS).  At
  $z\gsim$\,1 we include the cluster arc surveys of \citet{Swinbank06a}
  ($z$\,=\,1 arcs) and \citet{Jones10} ($z$\,=\,2--3 arcs) and at
  $z\gsim$2 we include the SINS and AMAZE surveys
  \citet{ForsterSchreiber09,Gnerucci11}.  The solid line denotes the
  $z$\,=\,0 $B$-band Tully-Fisher relation from \citep{Pierce92}.  {\it
    Right:} The evolution of the zero-point of the $B$-band
  Tully-Fisher relation.  The open symbols denote individual galaxies
  and the solid points show the median (and error) in seven redshift
  bins between $z$\,=\,0 and $z$\,=\,3.  The curves shows the
  predictions of the evolution of the $B$-band Tully-Fisher relation
  from semi-analytic models by \citet{Bower06} and \citet{Dutton11}
  which both predict evolution of $\Delta M_B<$\,1\,mags between
  $z$\,=\,0 and $z$\,=\,3, less than seen in the observations.}
\label{fig:TF_B}
\end{figure*}

In Fig.~\ref{fig:TF_B} we show the $z$\,=\,0 Tully-Fisher relation from
\citet{TullyFisher} and overlay the $z$\,=\,0.84--2.23 measurements
from our sample (using V$_{\rm 2.2}$ as the preferred measured of
rotational velocity).  We also include the field sample
($<z>$\,=\,0.33) from \citet{Bamford05}, the $z\sim$\,1 field samples
from \citet{Miller11} and \citet{Weiner06}, the $z$\,=\,1--3 lensed
samples from \citet{Swinbank06a} and \citet{Jones10} and the $z\sim$\,2
and $z\sim$\,3 star-forming galaxies from \citet{ForsterSchreiber09}
and \citet{Gnerucci11} respectively.

It is clear from Fig.~\ref{fig:TF_B} that the $z$\,=\,0 $B$-band
Tully-Fisher relation provides a poor fit to high-redshift data, with
most of the high-redshift galaxies lying above the local relation.
Since the high-redshift sample represents a heterogeneous mix of
populations, all with different and complex selection functions, we
limit our search for the evolution in the Tully-Fisher to its
zero-point only.  We therefore fix the slope to the $z$\,=\,0 relation
and in each of the seven redshift bins from $z$\,=\,0.0--3 we derive
the zero-point offset using a non-linear least squares regression
(weighted by the velocity errors in each galaxy) and show these results
in Fig.~\ref{fig:TF_B}.

For a fixed slope, the strongest evolution occurs between $z$\,=\,0--1,
and in this range the zero-point evolution follows $\Delta M_{\rm
  B}(mag)$\,=\,($-$1.9\,$\pm$\,0.3)\,$z$ up to $z\sim$\,1 and then
significantly flattens above $z$\,=\,1 to $\Delta$M$_{\rm
  B}(mag)$\,=\,($-$0.06\,$\pm$\,0.25)\,$z-$(1.68\,$\pm$\,0.25) up to
$z\sim$\,3.  Since the rest-frame $B$-band is dominated by recent
star-formation activity, the strong evolution in the rest-frame
$B$-band Tully-Fisher zero-point between $z$\,=\,0 and $z$\,=\,1
(2\,magnitudes or a factor $\sim$\,6 in luminosity) and then
flattening to $z\sim$3, is consistent with the evolution of the
star-formation volume density with redshift \citep{Hopkins06}.

We can compare this evolution to galaxy formation models, and in
Fig.~\ref{fig:TF_B} we also overlay the predicted zero-point offset
from the semi-analytic models from \citet{Dutton11} and
\citet{Bower06}.  The \citet{Dutton11} semi-analytic models suggest
that for a fixed circular velocity, the luminosity of a galaxy at
$z\sim$\,1 should be a factor up to $\sim$\,2.5\,$\times$ higher than
today \citep{Dutton11}, although the data are suggestive of markedly
stronger evolution than predicted by the models $z\sim$\,1.  In
contrast the predictions from the \citet{Bower06} model predicts very
little evolution in the $B$-band Tully-Fisher relation ($\Delta
M_B<$\,0.2 magnitudes between $z$\,=\,0--4).  In this model, the
apparent lack of evolution in the $B$-band Tully-Fisher relation in the
galaxy formation models occurs because the specific star-formation rate
evolves rapidly with both mass and redshift.  Although the model
approximately matches the evolution of the star-formation rate density
with redshift, this is offset by an increase in the specific
star-formation rate in low mass galaxies.  Moreover, since the dark
halos are more concentrated at high-redshift, a fixed v$_{circ}$ does
not select the same mass halo at the two epochs (for a fixed v$_{\rm
  circ}$, a halo at $z$\,=\,0 is a factor $\sim$1.7$\times$ more
massive than at $z$\,=\,2).  This combination of the evolution of the
specific star-formation rate with mass and redshift, and the evolution
in the halo densities for a fixed v$_{\rm circ}$ approximately cancel
to predict that there should be very little evolution in the $B$-band
Tully-Fisher relation.

\subsubsection{The Stellar Mass Tully-Fisher Relation}
\label{sec:STF}

\begin{figure*}
\centerline{\psfig{file=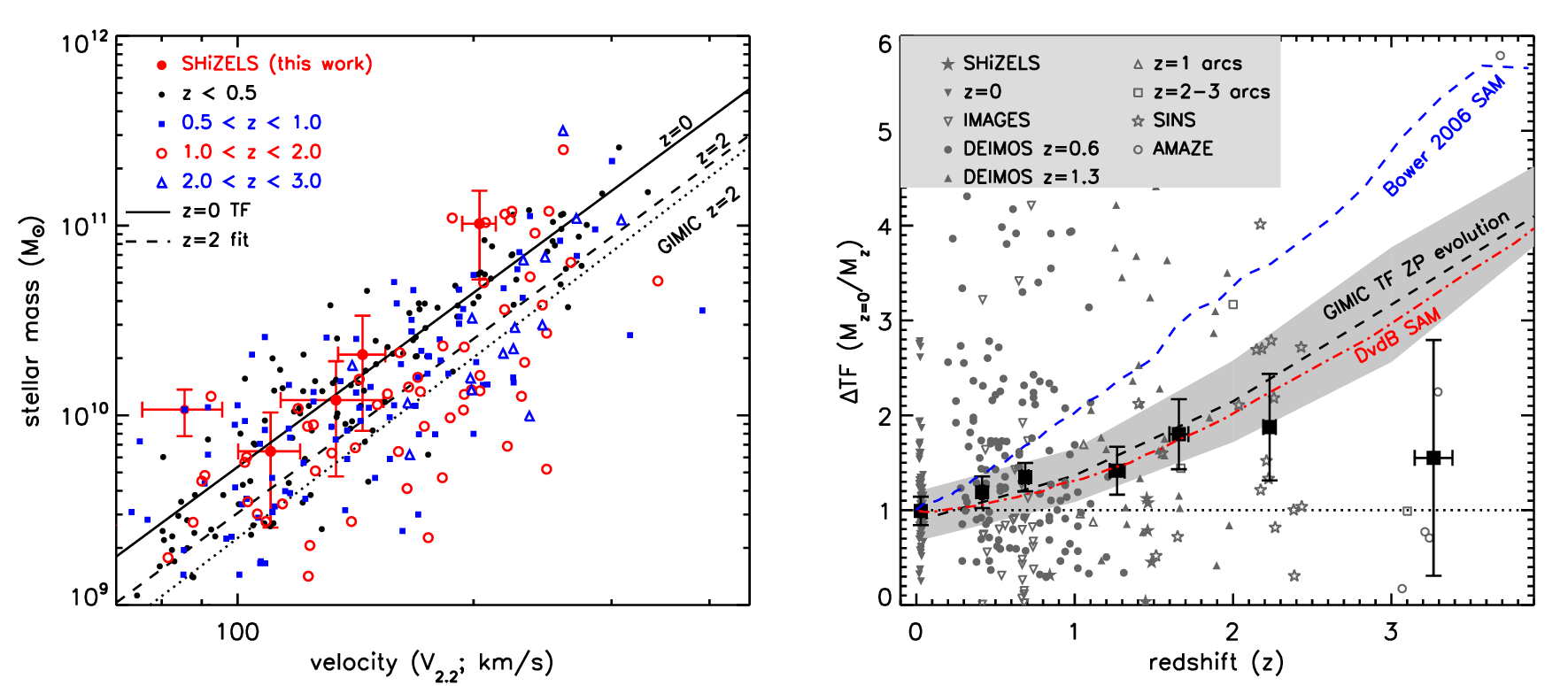,angle=90,width=7in}}
\caption{{\it Left:} The evolution of the stellar mass Tully-Fisher
  relation.  We baseline the evolution against the $z$\,=\,0 work from
  \citet{Pizagno05}.  The high-redshift points are compiled from the
  intermediate- and high- redshift ($z\sim$0.6 DEIMOS and $z\sim$1.3
  DEIMOS) observations from \citet{Miller11} and \citet{Miller12}; the
  $z$\,=\,1 and $z$\,=\,2--3 cluster arc surveys from
  \citet{Swinbank06a} and \citet{Jones10} with stellar masses from
  \citet{Richard11}; and the $z\sim$2--3.5 SINS and AMAZE surveys from
  \citet{Cresci09} and \citet{Gnerucci11}.  The symbols show individual
  galaxies.  The solid line denotes the correlation at $z$\,=\,0 from
  \citet{Pizagno05} (corrected to a Chabrier IMF).  The dashed line is
  best-fitting zero point to the $z$\,=\,2 sample galaxies (for a fixed
  $v_{2.2}$) which shows an offset of of
  $\Delta$M$_{\star,z=0}$\,/\,M$_{\star}$\,=\,2.0\,$\pm$\,0.4 between
  $z$\,=\,0 and $z$\,=\,2.5.  The dotted lines denote the Tully-Fisher
  at $z$\,=\,2 from the numerical simulations from \citet{Crain09} and
  \citep[see also][]{McCarthy12}, which predict evolution in both the
  zero-point and slope.  Here, we concentrate on the zero-point
  evolution, and note that the predicted evolution for a disk with
  circular velocity 100--200\,km\,s$^{-1}$ is an increase in stellar
  mass of a factor 1.5--3\,$\times$ between $z$\,=\,2 and $z$\,=\,0
  (equivalently, at high-redshift the maximum circular velocity is
  greater for the same stellar mass which may be consistent with the
  galaxies being compact at high-redshift and larger at low-redshift).
  {\it Right:} The evolution of the zero-point of the Tully-Fisher
  Relation.  The symbols denote individual points (coded by the
  survey), whilst the solid symbols denote the average in six redshift
  bins.  We also overlay the redshift evolution of the zero-point of
  the Tully-Fisher relation from the numerical model from
  \citet{Crain09} as well as the semi-analytic models from
  \citet{Bower06} and \citet{Dutton11} (DvdB).  These galaxy formation
  models predict evolution in the zero-point of the Tully-Fisher
  relation out to $z\sim$\,3 which is consistent with the observed
  trend given the large uncertainties in the latter.}
\label{fig:TF}
\end{figure*}

The rest-frame $B$-band Tully-Fisher relation is sensitive to recent
star formation, and so perhaps the more fundamental relation is the
stellar mass Tully-Fisher relation (M$_{\star}$ versus v$_{\rm asym}$)
which reflects the time-integrated star-formation history.  The
$z$\,=\,0 stellar mass Tully-Fisher relation is well established
\citep{Bell01,Pizagno07,Meyer08}, and in Fig.~\ref{fig:TF} we combine
the stellar mass estimates with the inclination-corrected rotation
speed of our galaxies to investigate how this varies as a function of
redshift.  Again, in this plot, we also include a number of
high-redshift measurements; at $z\sim$\,0.6 and $z\sim$\,1.3 from
\citet{Miller11} and \citep{Miller12}, the $z$\,=\,1--3 lensed galaxy
surveys \citep{Swinbank06a,Jones10,Richard11} as well as the the
$z\sim$\,2 SINS and $z>$\,3 SINFONI IFU surveys
\citep{Cresci09,Gnerucci11}.  Where necessary, we have corrected the
samples to use V$_{2.2}$ using V$_{2.2}$/V$_{max}$\,=\,0.93 if only
V$_{\rm max}$ is given in the literature (see also \cite{Dutton07}).

In comparison to the rest-frame $B$-band Tully-Fisher relation,
Fig.~\ref{fig:TF} shows that the the redshift evolution in the stellar
mass Tully-Fisher relation is weak.  Due to the heterogeneous selection
functions of these various surveys, we again restrict the search to the
evolution in the zero-point with redshift.  We therefore construct the
same seven redshift bins from $z$\,=\,0--3 as in \S~\ref{sec:BTF} and
measure the zero-point offset in each bin (using a non-linear least
squares regression weighted by the velocity errors in each galaxy;
(Fig.~\ref{fig:TF}).  The data suggest that the zero-point offset from
$z$\,=\,2.5 to $z$\,=\,0 is
$\Delta$M$_{*,z=0}$\,/\,M$_{\star}$\,=\,2.0\,$\pm$\,0.4 (i.e. at a
fixed circular velocity the stellar mass of galaxies has increased by a
factor $\sim$\,2.0 between $z$\,=\,2.5 and $z$\,=\,0).  The weak
evolution in the stellar mass Tully-Fisher relation (in particular
below $z\sim$\,1) is due to the fact that individual galaxies evolve
roughly along the scaling relation rather than due to weak evolution in
the galaxies themselves \citep[e.g.\ ][]{Portinari07,Brooks11}.  This
indicates that the baryon conversion efficiency,
$\eta$\,=\,(M$_{\star}$\,/\,M$_{\rm 200}$)\,/\,($\Omega_{\rm
  b}$\,/\,$\Omega_{\rm 0}$, is fixed with redshift.  


In Fig.~\ref{fig:TF} we also compare the zero-point evolution in the
Tully-Fisher relation to that from semi-analytic models from
\citet{Bower06} and \citet{Dutton11} and the cosmological hydrodynamic
Galaxies--Intergalactic Medium Interaction Calculation (GIMIC,
\citealt{Crain09}).  The \citet{Bower06} and GIMIC simulations both use
the same underlying dark matter distribution, although the GIMIC
simulation follows the hydrodynamical evolution of gas, radiative
cooling, star formation and chemo-dynamics, and includes a prescription
for energetic feedback from supernovae.  In contrast to simulations of
small periodic volumes, GIMICs unique initial conditions enable it to
follow, at high-resolution, cosmologically representative volumes to
$z$\,=\,0 \citep[see][]{Crain09,McCarthy12}.  To search for evolution
in the Tully-Fisher zero-point in the simulations, we only consider
galaxies with star-formation rates $>$1\,M$_{\odot}$\,yr$^{-1}$ (at all
redshifts) so that a reasonable comparison to observational data can be
made.  Both the semi-analytic and hydrodynamic simulations shown in
Fig.~\ref{fig:TF} predict strong evolution, with the GIMIC simulation
suggesting that, at a fixed circular velocity the stellar mass of
galaxies should increase by 2.5\,$\pm$\,0.5\,$\times$ between
$z$\,=\,2 and $z$\,=\,0.  This evolution is matched by the
semi-analytic model from \citet{Dutton11}.  However, the semi-analytic
model from \citet{Bower06} predicts somewhat stronger evolution in the
stellar mass for a fixed circular velocity.


It is interesting to note that the GIMIC simulation predicts that the
scatter in the Tully-Fisher relation should not evolve strongly with
redshift (for a fixed circular velocity, the scatter is 25\,$\pm$\,5\%
over the redshift range $z$\,=\,0--3).  Our dynamical measurements,
particularly when combined with those from \citet{Weiner06} and
\citet{Miller12}, are sufficiently large in number to compare to the
models, and Fig.~\ref{fig:TF} shows that the scatter in the
observations is consistent with the models (for all but the last bin in
redshift where the number of data-points is $<$8), , suggesting that the
samples may now be sufficiently large (with rotation curves well enough
sampled) that the scatter is intrinsic.



%
%
%
\begin{figure}
\centerline{\psfig{file=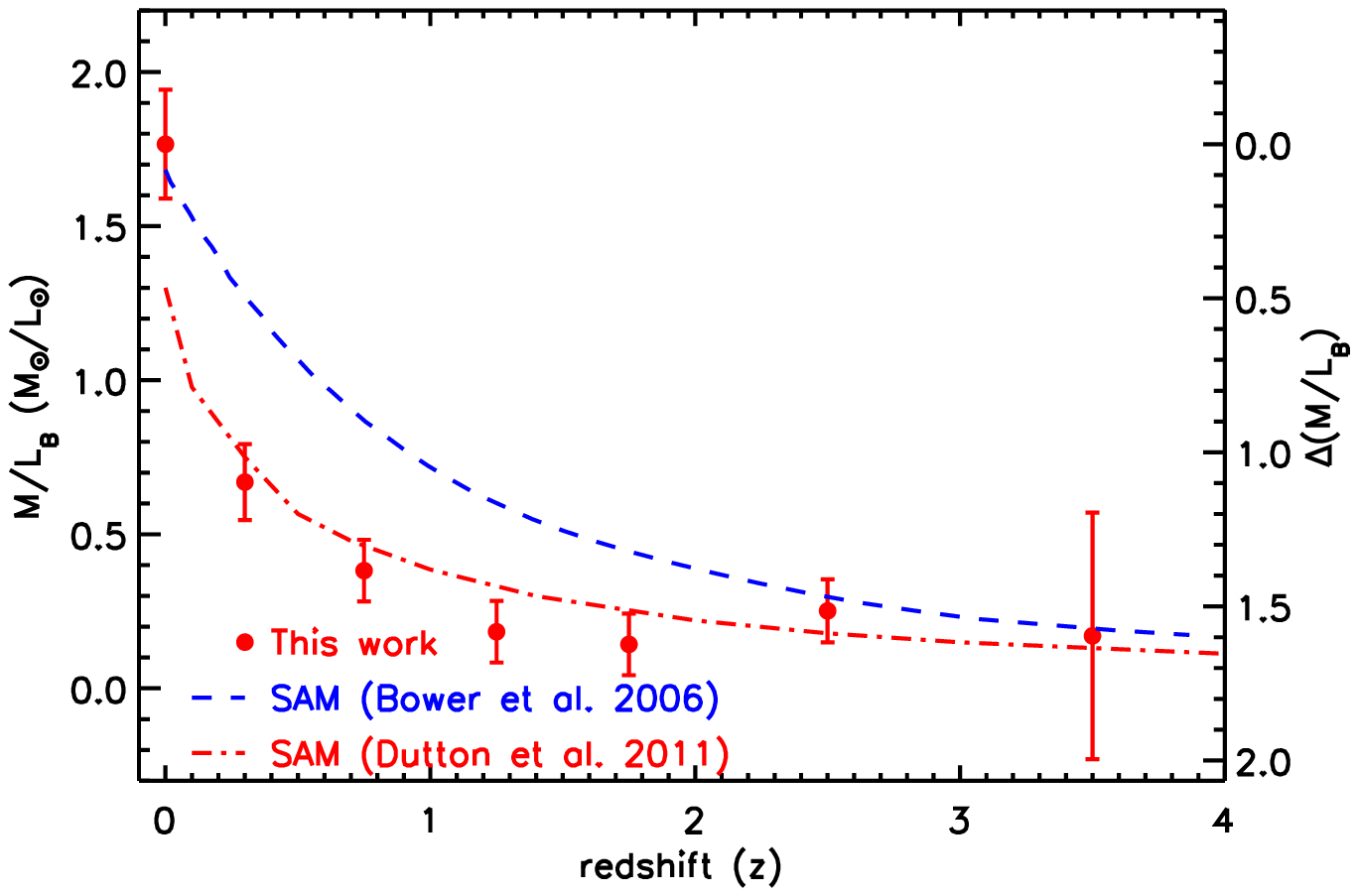,angle=90,width=3.5in}}
\caption{The $B$-band mass-to-light ratio as a function of redshift for
  star-forming galaxies from this work plus the compilation of studies
  shown in Fig.~\ref{fig:TF_B} and ~\ref{fig:TF}.  The dashed- and
  dot-dashed- lines show the predicted evolution of the mass-to-light
  ratio as a function of redshift from the \citet{Bower06} and
  \citet{Dutton11} semi-analytic models.  This figure shows that in
  both the observations and galaxy formation models there is strong
  evolution in the mass-to-light ratio of galaxies with redshift.  The
  strongest evolution in the data is seen for $z$\,=\,0--1, with
  significant flattening above $z$\,=\,1.}
\label{fig:MLz}
\end{figure}

\subsection{The Redshift Evolution of the Mass-to-Light ratio}

Another route to examine the evolution in the $B$-band and stellar mass
Tully-Fisher relations is to combine the offsets and measure the
evolution of the mean mass-to-light ratio with redshift.  We caution
that the conversion of zero-point offsets to offsets in mass-to-light
assumes that star-forming galaxies form a homologous family and that
the evolution is a manifestation of underlying relations between
mass-to-light ratio and other parameters, such as star-formation
history, gas accretion and stellar feedback.

Under these assumptions, in Fig.~\ref{fig:MLz} we show the evolution of
the rest-frame $B$-band mass-to-light ratio.  As expected from the
zero-point Tully-Fisher offsets, this figure shows that the average
mass to light ratio of star-forming galaxies decreases strongly from
$z$\,=\,0 up to $z$\,=\,1 and then flattens.  This behavior is
consistent with the previous demonstration that star-forming galaxies
at high-redshift have lower stellar masses and higher $B$-band
luminosities.  The strongest evolution occurs up to $z\sim$\,1,
$\Delta$M$_{\star}$\,/\,L$_{\rm B}$\,=\,1.1\,$\pm$\,0.2, consistent
with previous studies \citep{Miller11} (the fractional change in
mass-to-light ratio over this period is $\Delta$(M\,/\,L$_{\rm
  B}$)\,/\,(M\,/\,L$_{\rm B}$)$_{\rm z=0}\sim$\,3.5 between $z$\,=\,1.5
and $z$\,=\,0, with most of the evolution occuring below $z$\,=\,1).

We note that we examined whether the evolution in the mass-to-light
ratio could be reproduced using simple star-formation histories,
ranging from i) a constant star-formation rate (with a formation
redshift, $z_f$\,=\,4--8); or ii) a set of exponentially decreasing
star-formation rates with $e$-folding times ranging from 0.25--10\,Gyr
(and formation redshifts ranging from $z_f$\,=\,2--10).  However, using
these simple star-formation histories (adopting a Chabrier IMF with
0.5--1 solar metallicities and the Padova (1994) stellar evolution
tracks), we are unable to find a acceptable fit with a single
star-formation history.  This could be because the "average"
star-formation history is more complex than a simple star-formation
model, or because the current low and high-redshift data can not be
linked by a simple evolutionary model.  However, in Fig.~\ref{fig:MLz}
we also overlay the predicted evolution of the $B$-band mass-to-light
ratio from the semi-analytic models of \citet{Bower06} and
\citet{Dutton11}, both of which predict a sharp decline to $z\sim$\,1
and then flattening to higher redshift, which provides a reasonable
match to the observations.




\subsection{The Evolution of the Size--Rotational Velocity Relation}

In a $\Lambda$CDM cosmology, the sizes of galaxy disks and their
rotational velocities should be proportional to their parent dark
halos, and since the halos are denser at higher redshift, for fixed
circular velocity the disk sizes should scale inversely with Hubble
time \citep{Mo98}.  In this scenario, disks at $z$\,=\,1 and $z$\,=\,2
should be 1.8 and 3.0\,$\times$ smaller respectively at fixed circular
velocity.  With the high-resolution measurements for the galaxies in
our sample we can measure the extent of the H$\alpha$ (and together
with the dynamics) compare these relations at $z\sim$\,1.5 to
star-forming disks at $z$\,=\,0.

\begin{figure}
   \psfig{figure=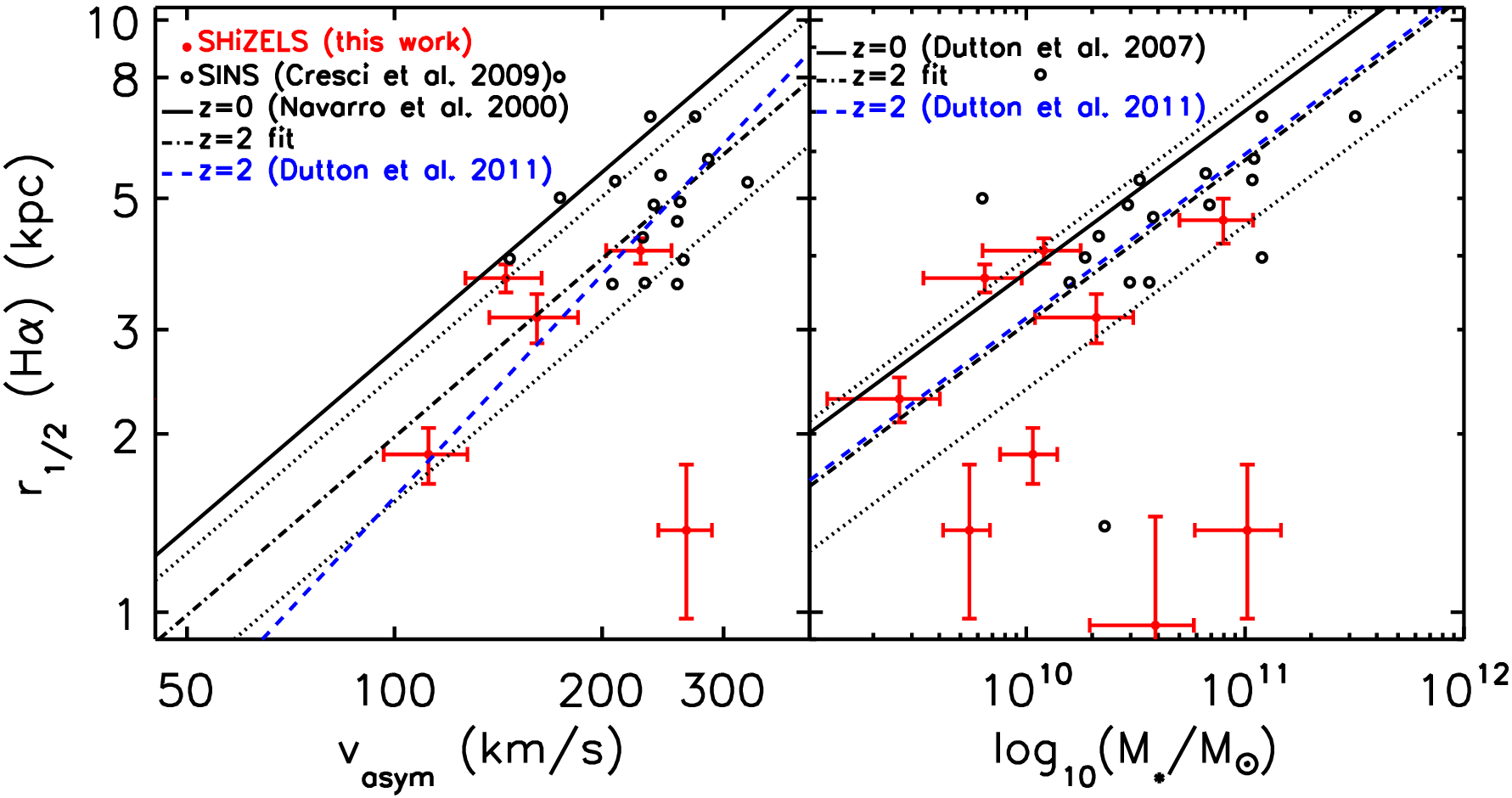,width=9cm,angle=90}
\caption{{\it Left:} The relation between the circular-velocity and
  H$\alpha$ half-light radius for high-redshift galaxies.  We include
  the results from the SHiZELS galaxies in our sample as well as those
  from SINS \citep{Cresci09}.  The solid line shows the $z$\,=\,0
  relation from \citet{Navarro00} (which is based on a compilation from
  \citealt{MathewsonFord92} and \citealt{Courteau97}).  Fitting the
  local relation (assuming a fixed slope) to the high-redshift points
  suggests and offset $\Delta r_{\rm h}$\,/\,r$_{\rm
    h,z=0}$\,=\,1.4\,$\pm$\,0.3, which is shown by the dot-dashed line
  (the scatter is shown by the dotted lines).  This evolution is
  consistent with that predicted by halo--disk scaling arguments
  (e.g.\ \citealt{Dutton11}; blue dashed line).  {\it Right:} the
  relation between stellar-mass and half-light radius in the SHiZELS
  and SINS surveys.  The solid line shows the $z$\,=\,0 relation from
  \citet{Dutton11} whilst the dot-dashed curve shows the offset of the
  high-redshift measurements from the local relation (assuming a fixed
  slope), which shows a $\Delta r_{\rm rh}$\,/\,r$_{\rm
    h}$\,=\,1.2\,$\pm$\,0.3 offset for a fixed stellar mass.  We also
  show the predicted evolution of $z$\,=\,2 galaxies from the $z$\,=\,0
  relation from the \citet{Dutton11} semi-analytic model.  }
\label{fig:rh_vmax}
\end{figure}

In Fig.~\ref{fig:rh_vmax} we show the relation between the H$\alpha$
half-light radius and rotational velocity for the galaxies in our
sample (using the asymptotic velocity as measured from the disk
modeling).  In this plot, we also include the measurements from the
SINS galaxies from \citet{Cresci09}.  First, we note that the H$\alpha$
half-light radii for the SHiZELS galaxies
($r_h$\,=\,2.4\,$\pm$\,0.8\,kpc) are comparable to the average
rest-frame UV/optical half light radii of star-forming galaxies at this
redshift \citep[e.g.][]{Conselice03a,Law07b,Swinbank10b}.

In order to compare to $z$\,=\,0 data, we show the measurements of
local disks from \citet{Navarro00} (which is based on a compilation
from \citealt{MathewsonFord92} and \citealt{Courteau97}), but
converting the disk scale length (r$_{\rm d}$) to half-light radius
using r$_{1/2}$\,=\,1.68\,r$_{\rm d}$, as appropriate for a disk with
an exponential profile.  As can be seen from this figure, the
high-redshift galaxies have half-light sizes (deconvolved for seeing)
which are smaller for fixed circular velocity than those at $z$\,=\,0.
Fitting the local the $z$\,=\,0 relation to the high-redshift data
(only allowing a zero-point shift and inverse weighting the fit by the
errors on each data-point) suggests that this offset is a factor
$\Delta r_{\rm h}$\,/\,r$_{\rm h,z=0}$\,=\,1.4\,$\pm$\,0.3, which is
consistent (given the smaller number statistics in the high-redshift
measurements) with that predicted from halo--disk scaling arguments
(\citealt{Mo98}; see also \citealt{Dutton11}).

The product of the rotational velocity and size of the disk define the
specific angular momentum, and the evolution (at a fixed circular
velocity) with redshift also encodes information on the relation
between the star-formation processes in the disk and the halo spin
parameters \citep[e.g.\ ][]{Navarro00,Tonini06}.  The offsets seen in
the r$_{1/2}$--v$_{\rm asym}$ relation compared to local scaling
relations in Fig.~\ref{fig:rh_vmax} translate directly into offsets in
the specific angular momentum for disks between $z$\,=\,0 and
$z\sim$\,2.  Indeed, the offsets in r$_{1/2}$--v$_{\rm asym}$ suggest
that, for fixed circular velocity the specific angular momentum of
disks at $z$\,=\,2 is a factor 1.4\,$\pm$\,0.3 lower than those at
$z$\,=\,0 \citep{Navarro00}.  Thus, if high-redshift disks evolve into
local disk galaxies, then the specific angular momentum must be
increased, either by removing material with low specific angular
momentum through outflows, or the galaxy disk must decouple from the
halo and acquire angular momentum from accreting gas.

Finally, in Fig.~\ref{fig:rh_vmax} we also show the relation between
stellar mass and H$\alpha$ half light radius for the galaxies in our
sample compared to the $z$\,=\,0 relation from \citet{Dutton07}.  The
high-redshift galaxies are offset from the local relation by $\Delta
r_{\rm rh}$\,/\,r$_{\rm h}$\,=\,1.2\,$\pm$\,0.3 for a fixed stellar
mass.  Although this is consistent with no evolution, we note that the
predicted evolution for a fixed stellar mass between $z$\,=\,0--2 from
the semi-analytic model from \citet{Dutton11} is $\Delta r_{\rm
  rh}$\,/\,r$_{\rm h}$\,=\,1.2.


%
%
%

\subsection{Metallicities and Spatially Resolved Chemical Abundances}
\label{sec:Zgrad}

The internal enrichment (and radial abundance gradients) of
high-redshift star-forming galaxies provides a tool for studying the
gas accretion and mass assembly process such as gas exchange
(inflows/outflows) with the intergalactic medium.  However, the small
sizes of high redshift galaxies (typically r$_{\rm h}\sim$\,2--3\,kpc;
e.g. \citealt{Law07b,Conselice08,Smail07,Swinbank10b}) mean that
sub-kpc resolution observations are required to measure the abundances
reliably.  At high redshifts the gas-phase metallicity of individual
galaxies can only be measured with limited accuracy, usually through
the [N{\sc ii}]\,/\,H$\alpha$ emission-line ratio as an indicator of
oxygen abundance and calibrations between gas phase emission line and
stellar absorption metallicities, such as those obtained by
\citet{Pettini04}.  Recently, the first reliable spatially resolved
measurements of [N{\sc ii}]\,/\,H$\alpha$ have been made in two highly
amplified galaxies at $z\sim$\,1.5--2.5, measuring strong, negative
gradients and suggesting that if $z\sim$\,2 starbursts evolve into
spiral disks today, then the gradients must flatten by
$\Delta$\,log(O\,/\,H)\,/\,$\Delta$R$\sim$\,0.05--0.1\,dex\,kpc over
the last 8--10\,Gyr \citep{Jones10b,Yuan11}.  Other studies at lower
resolution have derived weaker gradients that are negative, or
consistent with zero for rotating and interacting galaxies
\citep{Cresci10,Queyrel12}.  Positive gradients for interacting systems
may be a signature of the redistribution of the metal-rich gas produced
from a central starburst; \citep{Werk10,Rupke10}, or possibly as a
result of inflow of relatively unenriched gas from the halo (or IGM) if
it is able to intersect the central regions of the disk galaxy without
being disrupted.

\begin{figure}
   \psfig{figure=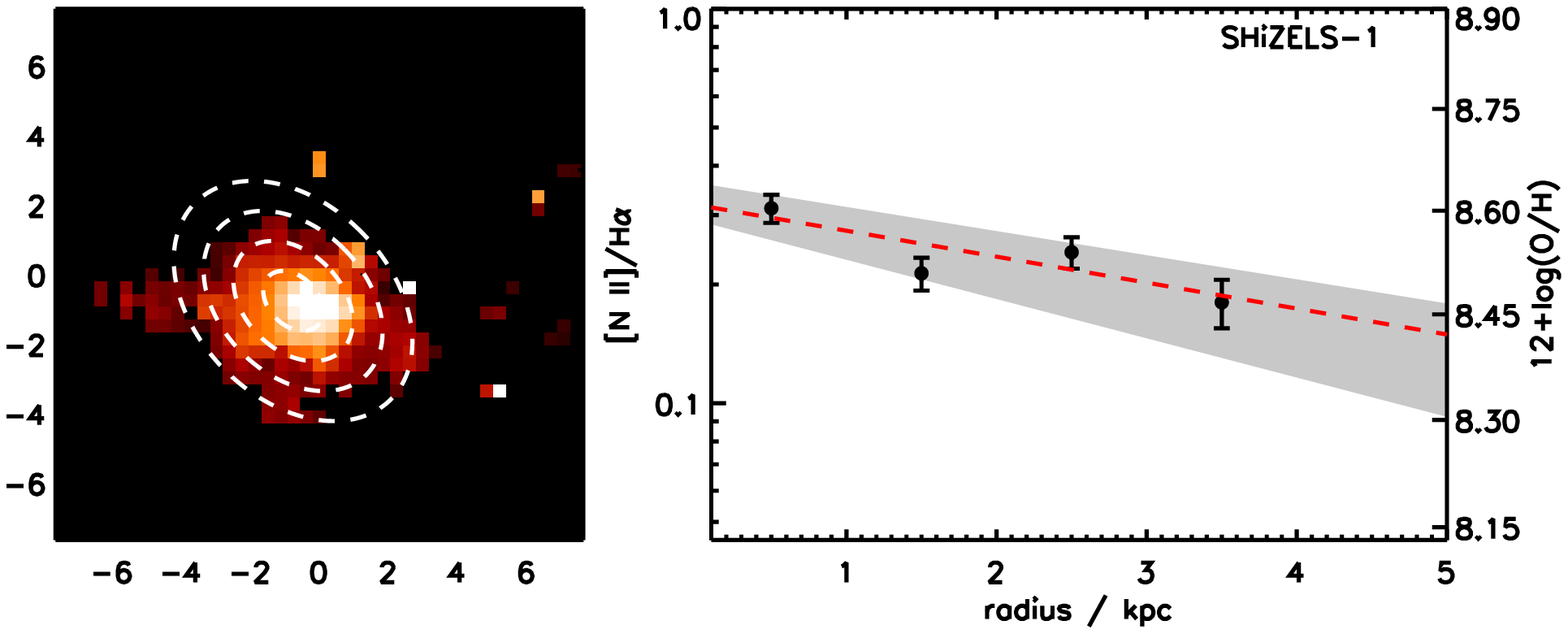,width=7.6cm,angle=0}
   \psfig{figure=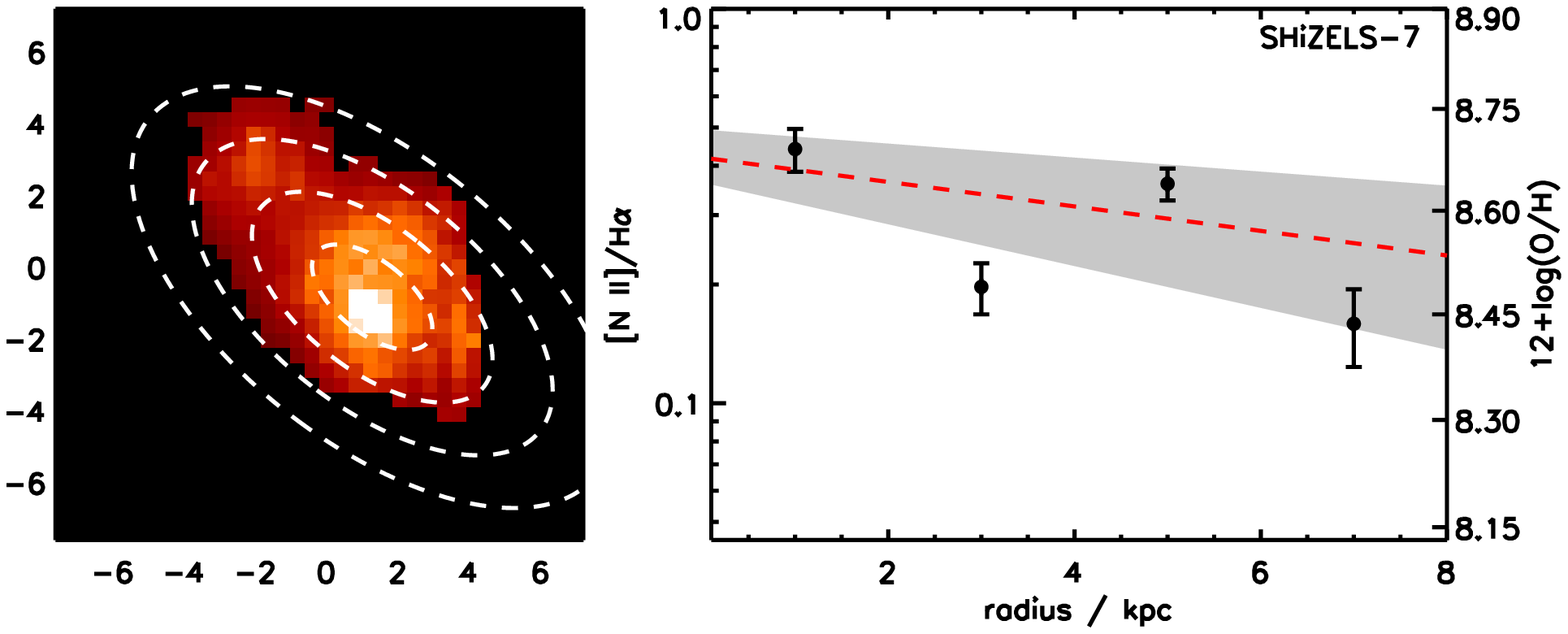,width=7.6cm,angle=0}
   \psfig{figure=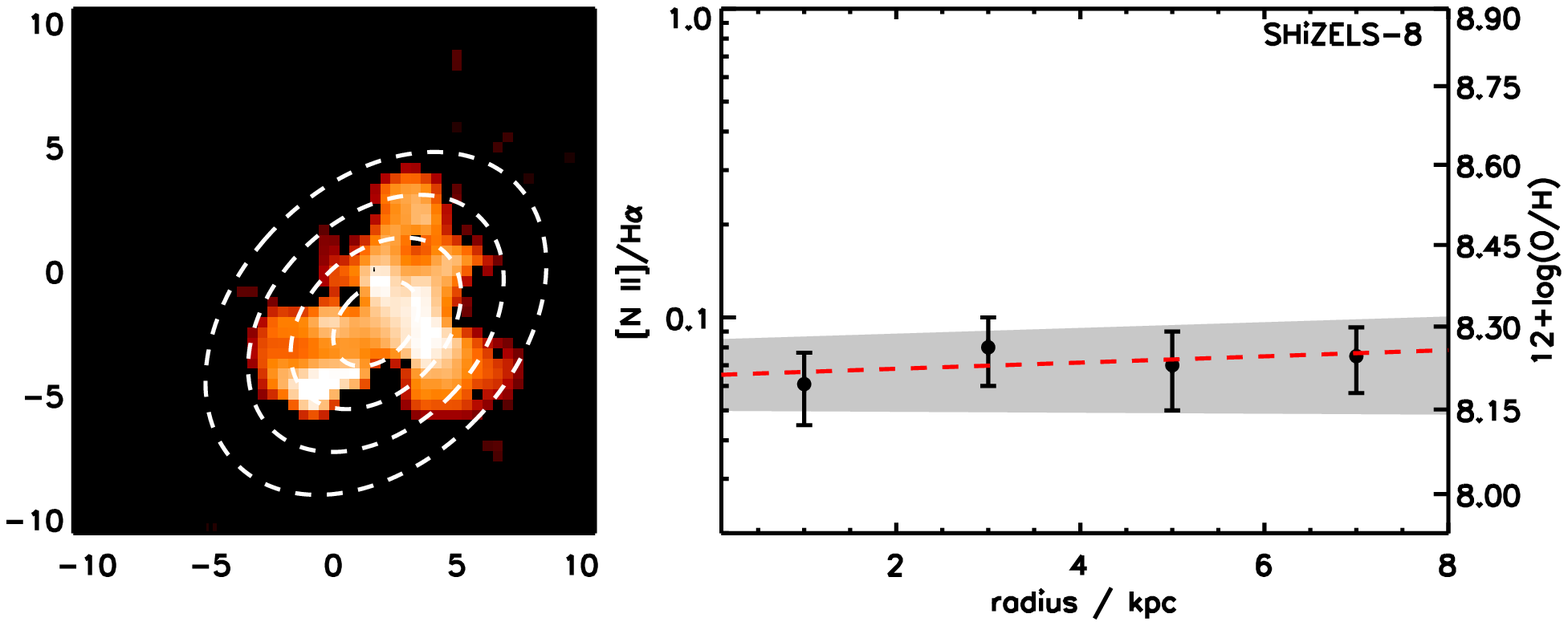,width=7.6cm,angle=0}
   \psfig{figure=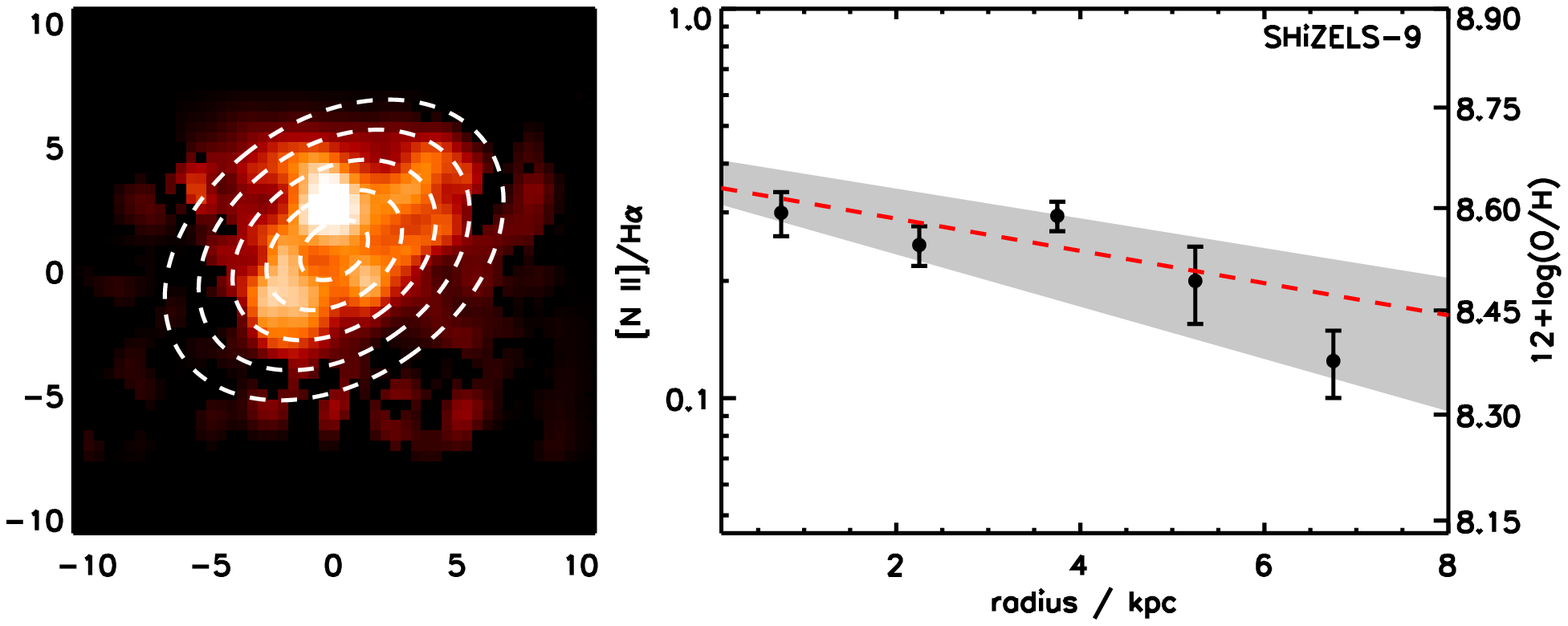,width=7.6cm,angle=0}
   \psfig{figure=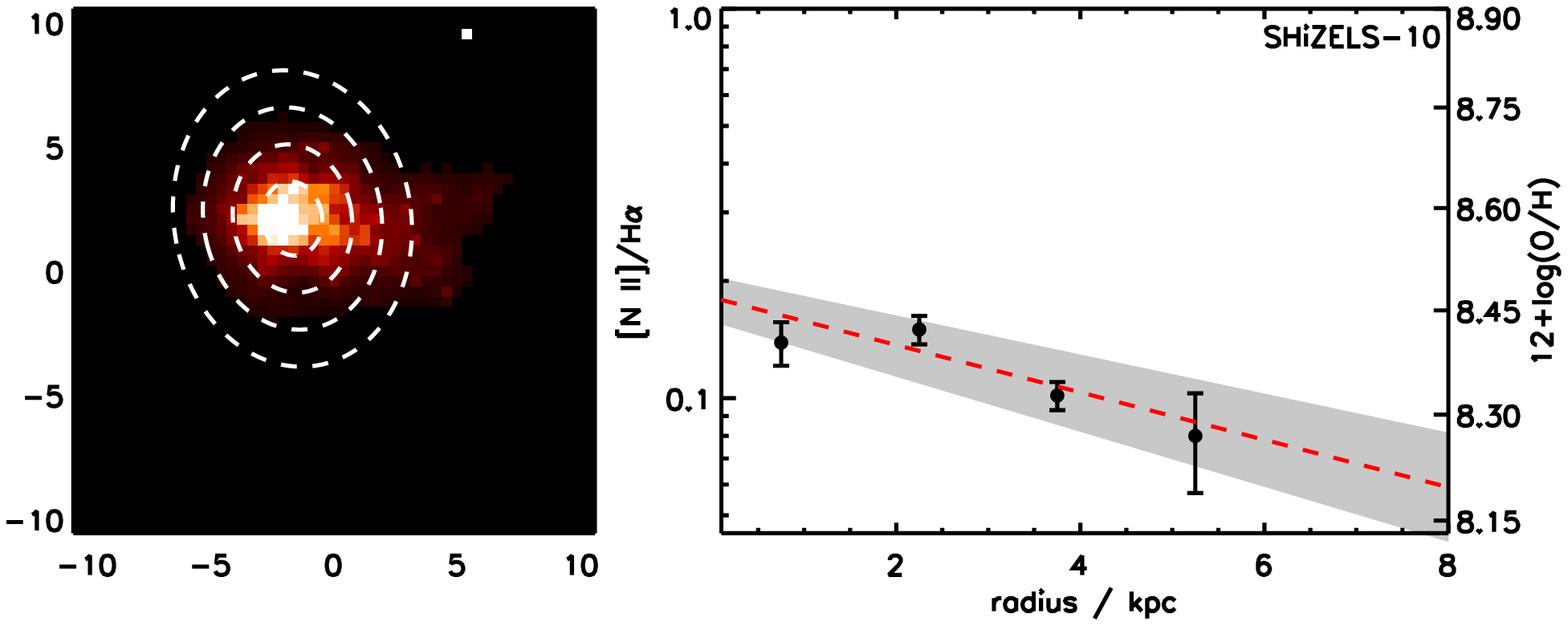,width=7.6cm,angle=0}
   \psfig{figure=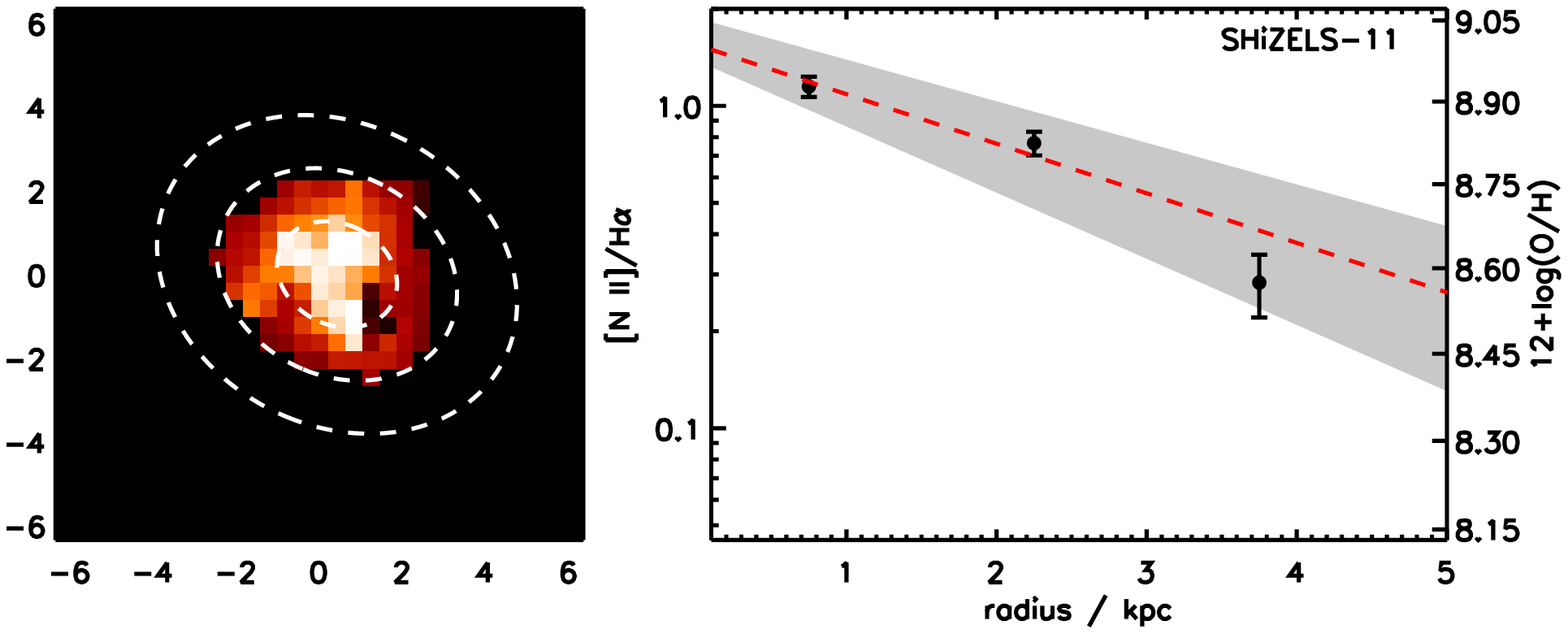,width=7.6cm,angle=0}
   \psfig{figure=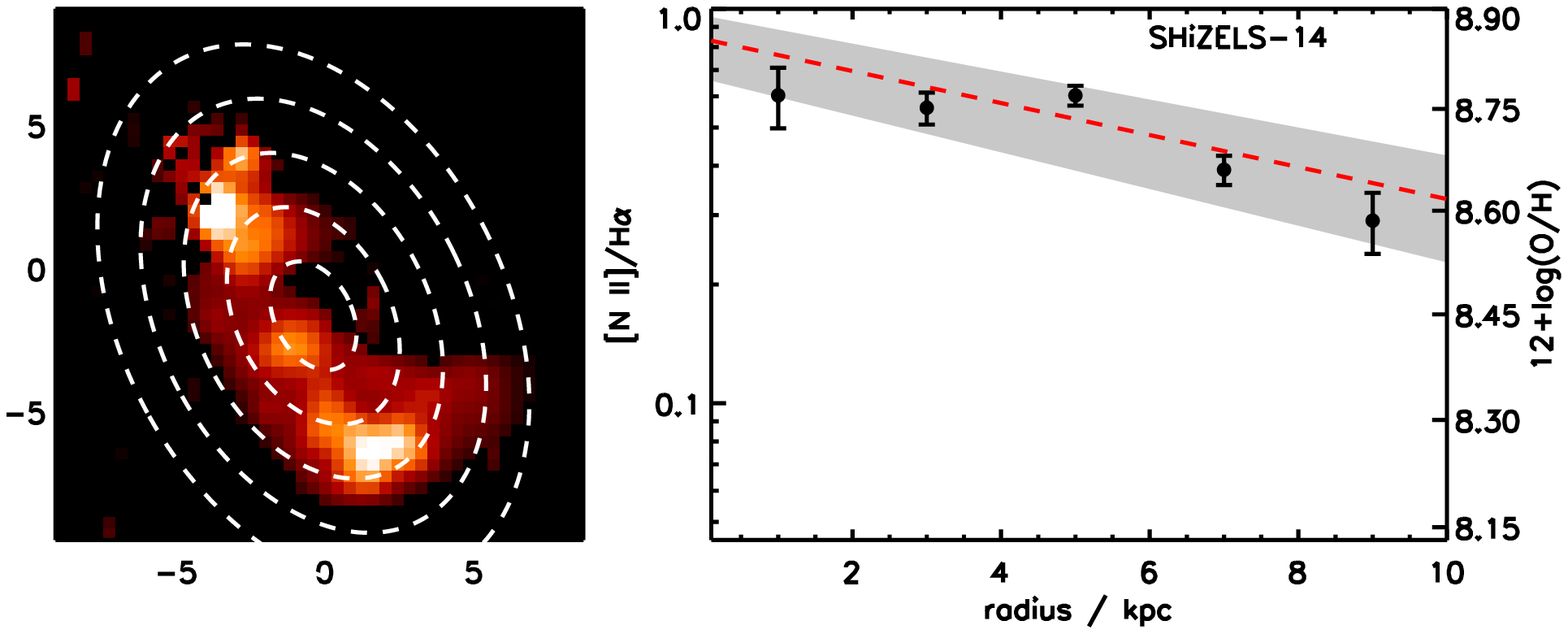,width=7.6cm,angle=0}
\caption{Metallicity gradients for the seven galaxies in our sample
  where spatially resolved measurements can be made. {\it Left:}
  Velocity integrated H$\alpha$ emission line maps.  The dashed annuli
  denote the $\sim$\,1\,kpc regions from which the [N{\sc
      ii}]/H$\alpha$ were extracted.  {\it Right:} The [N{\sc
      ii}]$\lambda$\,6583\,/\,H$\alpha$ emission line ratio as a
  function of physical radius.  We overlay the best-fit linear
  regression and we report the gradients for individual galaxies in
  Table~1.  In all cases the gradients are negative or consistent with
  zero, as expected for inside-out growth.}
\label{fig:shizels_gradient}
\end{figure}

\begin{figure}
   \psfig{figure=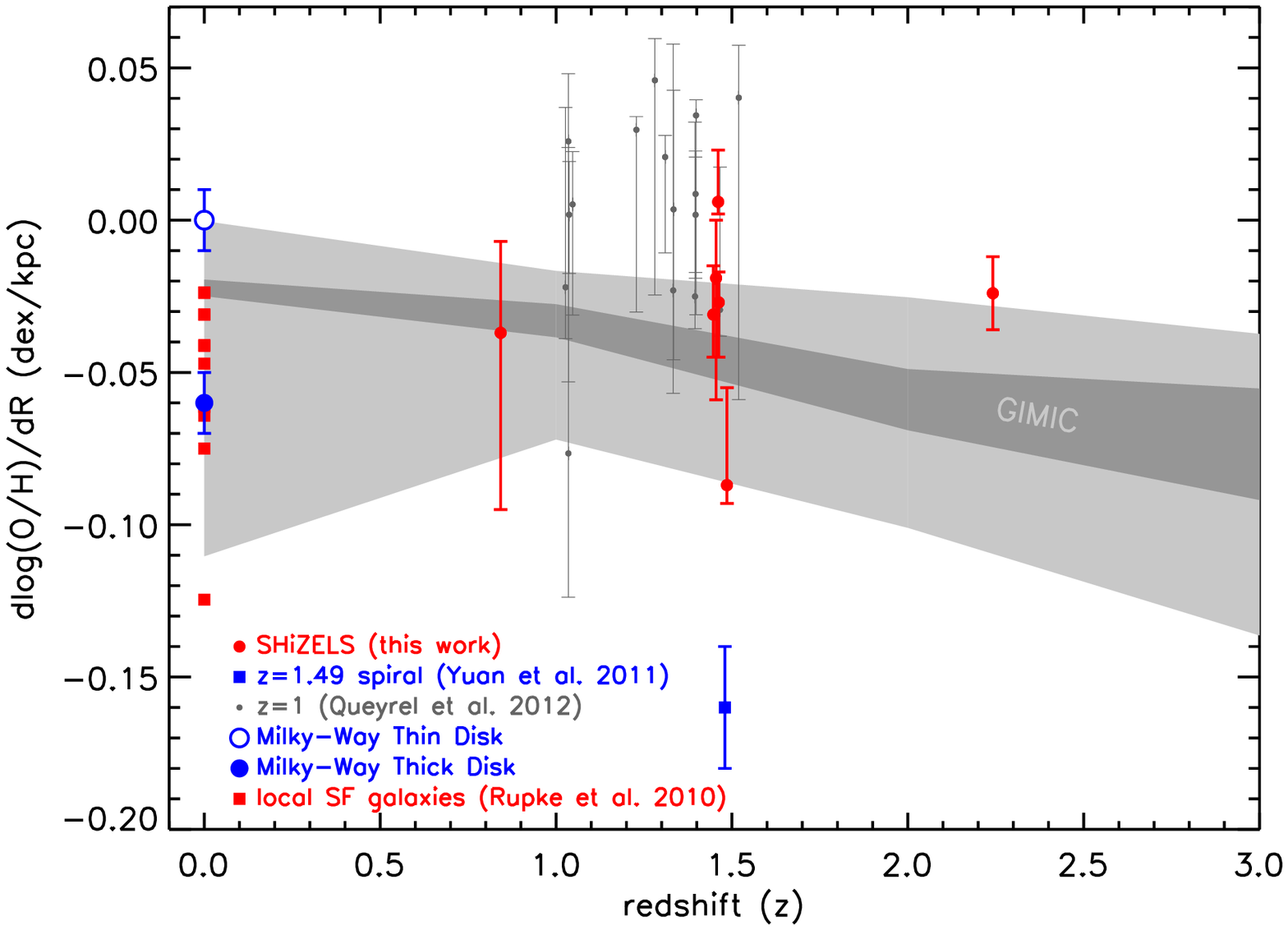,width=3.5in,angle=0}
\caption{Metallicity gradients versus redshift for the galaxies in our
  sample compared to the Milky-Way and other low- and high-redshift
  observations.  The Milky-Way thin and thick disk measurements are
  from \citet{Gilmore95,Bell96,Robin96,Edvardsson93}; the low-redshift
  star-forming galaxies are from \citet{Rupke10} and the $z$\,=\,1--1.5
  data from \citet{Yuan11} and \citet{Queyrel12}.  The median abundance
  gradient for our SHiZELS sample is
  $\Delta$\,log(O\,/\,H)\,/\,$\Delta$R\,=\,$-$0.027\,$\pm$\,0.005\,dex\,kpc$^{-1}$
  which is consistent with the thick disk of the Milky-Way.  We also
  show the theoretical evolution of the metallicity gradient with
  redshift from the GIMIC simulation \citep{Crain09,McCarthy12}.  The
  dark grey shows the range of metallicity gradients for all disk
  galaxies in the simulation in the mass range
  9.5\,$<$\,log(M\,/\,M$_{\odot}$)\,$<$\,11.5, whilst the light grey
  denotes the 1$\sigma$ scatter at each epoch.  The GIMIC simulation
  predicts the gradients should be steep around $z\sim$\,3 where gas
  accretion from the IGM is at its most efficient, but should flatten
  by a factor $\sim$2\,$\times$ between $z$\,=\,3 and $z$\,=\,0.}
\label{fig:Zgrad}
\end{figure}

With our high-resolution (and high signal-to-noise) resolved
observations we can measure the spatially resolved emission line ratios
and hence the distribution of chemical abundances.  First, we note that
the average galaxy integrated [N{\sc ii}]\,/\,H$\alpha$\, ratio for the
galaxies in our sample is [N{\sc ii}]\,/\,H$\alpha$\,
=\,0.27\,$\pm$\,0.09, which corresponds to
12\,+\,log(O\,/\,H)\,=\,8.58\,$\pm$\,0.07 (equivalently,
$\sim$\,0.73\,$\pm$\,0.16\,Z$_{\odot}$ where Z$_{\odot}$ is
12\,+\,log(O\,/\,H)\,=\,8.66\,$\pm$\,0.05; \citealt{Asplund04}).  We
caution that the [N{\sc ii}]/H$\alpha$ ratio is sensitive to shock
excitation, the ionization state of the gas, and the hard ionization
radiation field from an active galactic nucleus
\citep{Kewley02,Kewley06,Rich11}.  However, our galaxy-integrated
      [N{\sc ii}]\,/\,H$\alpha$ ratios are consistent with star
      formation throughout the disk (Fig.~\ref{fig:1dspec}).

Of the nine galaxies in our high-redshift sample, seven have [N{\sc
    ii}]$\lambda$6583 emission which is sufficiently strong (S/N$>$10
in the combined spectra) to allow spatially resolved studies of the
chemical abundance gradients.  First we subtract the large-scale
velocity motions from the data-cube (as defined by the best-fit
dynamical model).  We then use the disk inclination and position angle
to define $\sim$1\,kpc annuli (centered at the dynamical center) and
integrate the spectrum in each annulus to measure the [N{\sc
    ii}]\,/\,H$\alpha$ ratio.  In Fig.~\ref{fig:shizels_gradient} we
show the H$\alpha$ emission line maps as well as the spatially resolved
[N{\sc ii}]\,/\,H$\alpha$ ratio and overlay the best-fit linear
regression in each case.  We report the gradients individually in
Table~2.  In all cases, the metallicity gradients are negative or
consistent with zero, with an average
$\Delta$\,log(O\,/\,H)\,/\,$\Delta$R\,=\,$-$0.027\,$\pm$\,0.005\,dex\,kpc$^{-1}$,
and similar to those recently found for a sample of $z\sim$\,1
starbursts \citep{Queyrel12}.  This average gradient is steeper, but
consistent with the thick disk of the Milky-Way,
$\Delta$\,log(O\,/\,H)\,/\,$\Delta$R\,=\,$-$0.01\,$\pm$\,0.02\,dex\,kpc$^{-1}$
\citep{Gilmore95,Bell96,Robin96,Edvardsson93}, which has a mass
weighted age of 8--12\,Gyr (equivalently, a formation redshift
$z$\,=\,2--4; \citealt{Bensby11,Ruchti11}).



In Fig.~\ref{fig:Zgrad} we compare the metallicity gradients to those
from the GIMIC simulation (which include cold gas accretion from the
IGM).  We calculate the disc phase metallicity in cylindrical
coordinates (along the disc) at $z$\,=\,0, 1, 2 \& 3 in radial bins of
width 1\,kpc for all galaxies in the simulation with masses between
M$_{\star}$\,=\,10$^{9.5-11}$\,M$_{\odot}$.  At $z\sim$\,2, the model
disk galaxies from the GIMIC simulation display negative metallicity
gradients which are consistent with our observations.  Moreover, between
$z$\,=\,2 and $z$\,=\,0 the GIMIC simulation predicts that the gradient
should flatten by a factor $\sim$\,2.  In the simulation, this is a
consequence of the outer disk enrichment increasing with decreasing
redshift (as the gas inflow rates decline and gas is redistributed)
progressively flattening the radial gradient within the thick disk.
This is qualitatively similar to the numerical models of
\citet{Dekel09} and where the ``cold streams'' (which are most
prevalent around $z\sim$2) penetrate the halo and intersect and deposit
relatively unenriched gas at onto the galaxy disks at $\sim$15\,kpc
\citep{Dekel09,Dekel09b}.

\section{Conclusions}
\label{sec:conc}

We have presented AO-assisted, spatially resolved spectroscopy of nine
star-forming galaxies at $z$\,=\,0.84--2.23 selected from the UKIRT,
wide-field narrow-band, HiZELS survey \citep{Sobral12b}.  Our main
results can be summarised as:

$~\bullet$ Using H$\alpha$ emission line dynamics, we find that the
ratio of dynamical-to-dispersion support for the sample is $v_{\rm
  max}$\,sin($i$)\,/\,$\sigma$\,=\,0.3--3, with a median of
1.1\,$\pm$\,0.3, which is consistent with similar measurements for both
AO and non-AO studies of star-forming galaxies at this
epoch\citep[e.g.\ ][]{ForsterSchreiber09}.  At least six galaxies have
dynamics that suggest that their ionised gas is in a large, rotating
disk (in at least four of these we detect the turn over in their
rotation curves).  For these galaxies that resemble rotating systems,
we model the galaxy velocity field and derive the
inclination-corrected, asymptotic velocity.

$~\bullet$ We combine our dynamical observations with a number of
previous studies of intermediate- and high-redshift galaxies (in
particular from
\citealt{Bamford05,Weiner06,ForsterSchreiber09,Jones10,Miller11} and
\citealt{Miller12}) to investigate the evolution of the Tully-Fisher
relation.  We show that there is strong evolution in the zero-point of
the rest-frame $B$-band Tully-Fisher relations with redshift such that
at a fixed circular velocity, the $B$-band luminosity increases by
$\Delta$\,M$_{\rm B}$\,=\,2\,mags up to $z\sim$\,2 (a change in
luminosity of a factor $\sim$\,6; e.g.\ see also
\citealt{Bamford05,Weiner06,Miller11}).

$~\bullet$ In contrast, the stellar mass Tully-Fisher relation shows
much more modest evolution; for a fixed circular velocity the average
stellar masses has increased by a factor 2.0$\pm$0.4 between $z$\,=\,2
and $z$\,=\,0 (e.g.\ see also \citealt{Cresci09,Miller12}).  The weak
evolution in the stellar mass Tully-Fisher relation (in particular
below $z\sim$\,1) is due to the fact that individual galaxies evolve
along the scaling relations rather than due to weak evolution in the
galaxies themselves \citep[e.g.\ ][]{Portinari07,Brooks11,Dutton11},
and indicates that the baryon conversion efficiency,
$\eta$\,=\,(M$_{*}$\,/\,M$_{\rm 200}$)\,/\,($\Omega_{\rm
  b}$\,/\,$\Omega_{\rm 0}$), is fixed with redshift.  The observations
also suggest that the scatter in high-redshift Tully-Fisher relation is
comparable to that predicted by the models, suggesting that the samples
may now be sufficiently large (with rotation curves well enough
sampled) that the scatter is intrinsic.

$~\bullet$ Combining the rest-frame $B$-band and stellar mass
Tully-Fisher relations we show that the mass-to-light ratio of
star-forming galaxies evolves strong with redshift, decreasing by
$\Delta$\,M\,/\,L$_{\rm B}$\,=\,1.1\,$\pm$\,0.2 between $z$\,=\,0 and
$z$\,=\,1, but then ceases to evolve further above $z$\,=\,1.  This
change in mass-to-light ratio in the $B$-band is a factor
$\Delta$\,(M\,/\,L$_{\rm B}$)\,/\,(M\,/\,L$_{\rm B}$)$_{\rm
  z=0}\sim$\,3.5.  We show that the evolution in the Tully-Fisher
relations and mass-to-light ratio is in line with both
cosmologically-based hydrodynamic simulations and semi-analytic models
of galaxy formation.

$~\bullet$ Using the H$\alpha$ half light radius and circular velocity,
we also find that the SHiZELS galaxies are a factor
1.5\,$\pm$\,0.3\,$\times$ smaller for fixed circular velocity than for
disks at $z$\,=\,0.  If the $z\sim$\,2 and $z$\,=\,0 disks are linked
in a simple evolutionary scenario then high-redshift disks must acquire
specific angular momentum, either by removing material with low
specific angular momentum (e.g.\ through outflows), or through
acquisition of angular momentum from accreted gas at late times.

$~\bullet$ Finally, we measure the spatially resolved [N{\sc
    ii}]\,/\,H$\alpha$ to measure the metallicity gradients.  In all
cases, the metallicity gradients are negative or consistent with zero,
with an average
$\Delta$\,log(O\,/\,H)\,/\,$\Delta$\,R\,=\,$-$0.027\,$\pm$\,0.005\,dex\,kpc$^{-1}$,
which is consistent with that seen in the thick disk of the Milky-Way.
We show that these metallicity gradients are comparable to those
predicted for the gas disks of star-forming galaxies in the
cosmologically based hydrodynamic simulations.  In these models, the
inner disk undergoes initial rapid collapse and star-formation with gas
accretion (either from the halo and/or IGM) depositing relatively
unenriched material at the outer disk, causing negative abundance
gradients.  At lower redshift (when gas accretion from the IGM is less
efficient) the abundance gradients flatten as the outer disk becomes
enriched by star-formation (and/or the redistribution of gas from the
inner disk).

Overall, we demonstrate that well-sampled, dynamical measurements of
high-redshift star-forming galaxies can constrain the rotation curves,
measure the evolution of the basic disk scaling relations and their
radial abundance gradients.  We show that the current observational
samples with well resolved data are now approaching sufficiently large
samples that it is possible to refine or refute models of disk galaxy
formation.


\section*{acknowledgments}

We would like to thank the anonymous referee for their constructive
report which significantly improved the content and clarity of this
paper.  We thank Mario van der Ancker for help and support with the
SINFONI planning/observations, and Richard Bower, Natascha
F\"orster-Schreiber, Phil Hopkins and John Helly for a number of very
useful discussions.  AMS gratefully acknowledges an STFC Advanced
Fellowship.  DS is supported by a NOVA fellowship.  IRS acknowledges
support from STFC and a Leverhume Senior Fellowship and TT acknoledges
support from the National Science Foundation under grant number NSF
PHY11-25915.  The data presented are based on observations with the
SINFONI spectrograph on the ESO/VLT under program 084.B-0300.


\end{document}